# Ultrafast dynamics of local charge order in a THz-induced metastable quantum state

Luis E. Parra López,[1*] Jingkai Quan,[2] Alkisti Vaitsi,[1] Vivien Sleziona,[1] Fabian Schulz,[1#] Angel Rubio,[2,3] Martin Wolf,[1] and Melanie Müller[1†*]

[1]*Department of Physical Chemistry, Fritz Haber Institute of the Max Planck Society, 14195 Berlin, Germany*
[2]*Max Planck Institute for the Structure and Dynamics of Matter, 22761 Hamburg, Germany*
[3]*Initiative for Computational Catalysis, Flatiron Institute, Simons Foundation, New York City, NY 10010, USA*

**Corresponding authors: lopez@fhi-berlin.mpg.de, m.mueller@fhi-berlin.mpg.de*

**Controlling quantum materials with ultrafast light pulses enables access to transient and metastable states that are inaccessible under equilibrium conditions. Yet their local dynamics remain poorly understood due to the challenge of resolving ultrafast processes with angstrom-scale spatial resolution. Here, we use terahertz scanning tunnelling microscopy (THz-STM) to probe coherent collective dynamics within a THz-induced metastable state in the layered charge density wave (CDW) material 1T-$TaS_2$. Following ultrafast photoexcitation, we locally resolve coherent oscillations of the CDW amplitude mode at 2.5 THz together with two previously unreported modes at 1.3 THz and 0.7 THz. Comparison with phonon calculations identifies these as interlayer breathing and shear vibrations that are sensitive to the stacking configuration. These coherent dynamics are observed within a THz-induced metastable state that exhibits long-lived and spatially inhomogeneous modifications of the local density of states within the insulating gap, while higher THz fields drive a local redistribution and disordering of Star-of-David clusters near defects and domain boundaries. Our results suggest that the THz-induced metastable state involves a modification of the local interlayer stacking configuration, and demonstrate the role of interlayer degrees of freedom in the ultrafast dynamics of light-induced phases in layered quantum materials.**

## INTRODUCTION

Ultrashort light pulses offer powerful means for controlling and probing emergent phases in quantum materials, providing access to electronic and structural properties far from equilibrium.[1] Examples include photoinduced phase transitions, ultrafast switching and coherent control of nonequilibrium states,[2–4] light-induced superconductivity,[5,6] Floquet-engineering of topological phases,[7–10] or the creation of metastable states with no equilibrium counterpart.[11,12] A central challenge in this field is to understand how the structural and electronic states of such phases evolve locally near defects and domain boundaries, where new electronic properties may emerge and new phases can nucleate. While pump-probe techniques provide key insights into their ultrafast response, they lack the spatial resolution to capture the emergence and dynamical evolution of nonequilibrium phases at the atomic scale.

Recent advances in coupling femtosecond optical or Terahertz (THz) pulses to scanning tunnelling microscopes (STM)[13,14] offer new possibilities to resolve ultrafast dynamics with atomic spatial resolution. In particular, THz-lightwave-driven STM (THz-STM) has emerged as a powerful tool for probing the dynamics of free carriers and quasiparticles,[15–19] the vibration of single molecules,[18,20,21] and coherent collective modes[22–24] on the sub-nanometre scale. However, applying THz-STM to correlated quantum materials with fragile ground states remains an outstanding challenge, as the intense THz near-fields can simultaneously probe and perturb their local electronic and structural properties,[25,26] complicating signal analysis while providing access to new light-induced phases.

Here, we use THz-STM to study the emergence and photoinduced dynamics of a THz-induced metastable state in the layered quantum material 1T-$TaS_2$ at the angstrom scale. 1T-$TaS_2$ hosts an insulating commensurate charge density wave (C-CDW) phase at low temperatures,[27–29] whose electronic properties are governed by the interplay of intralayer electron correlations and interlayer hybridization of Star-of-David (SoD) orbitals. While a single 1T-$TaS_2$ layer is driven into a Mott-insulating state due to intralayer electron correlations,[29–31] the insulating properties of bulk 1T-$TaS_2$ are highly sensitive to the interlayer stacking order,[32–37] doping[38–40] and strain.[41] Moreover, the electronic structure can be highly spatially inhomogeneous, with transitions between band- and Mott-dominated insulating behaviour occurring on nanometre scales.[34] Furthermore, single ultrafast laser pulses,[3,11,42] electrical pulses,[43,44] STM voltage pulses,[45,46] or surface defects[47] can modify the interlayer stacking and induce a long-lived metastable metallic mosaic phase that is composed of nanoscale domains. However, despite extensive research, the role of atomic-scale disorder, heterogeneity and defects in the formation of nonequilibrium phases and their ultrafast dynamics remains largely unexplored.

By combining THz-STM with ultrafast photoexcitation, we resolve femtosecond coherent oscillations of the local charge order in 1T-$TaS_2$ with a spatial resolution of individual CDW superlattice sites. Together with the CDW amplitude mode, we detect two low-frequency modes originating from coherent interlayer shear and breathing vibrations. We further show that these modes are measured within a THz-induced metastable state, which we associate with a change in the local interlayer orbital hybridization. By tracking the real-time modulation of the tunnelling conductance on millisecond timescales and by imaging THz-induced changes in the local CDW order and local density of states (LDOS), we directly monitor how the tip-enhanced THz pulses drive the 1T-$TaS_2$ surface into localized and spatially inhomogeneous metastable state. Our work highlights the emerging capabilities of THz-STM for exploring the spatiotemporal evolution of light-induced phases at the angstrom scale and across different time scales.

## RESULTS

### STM and THz-STM of 1T-$TaS_2$

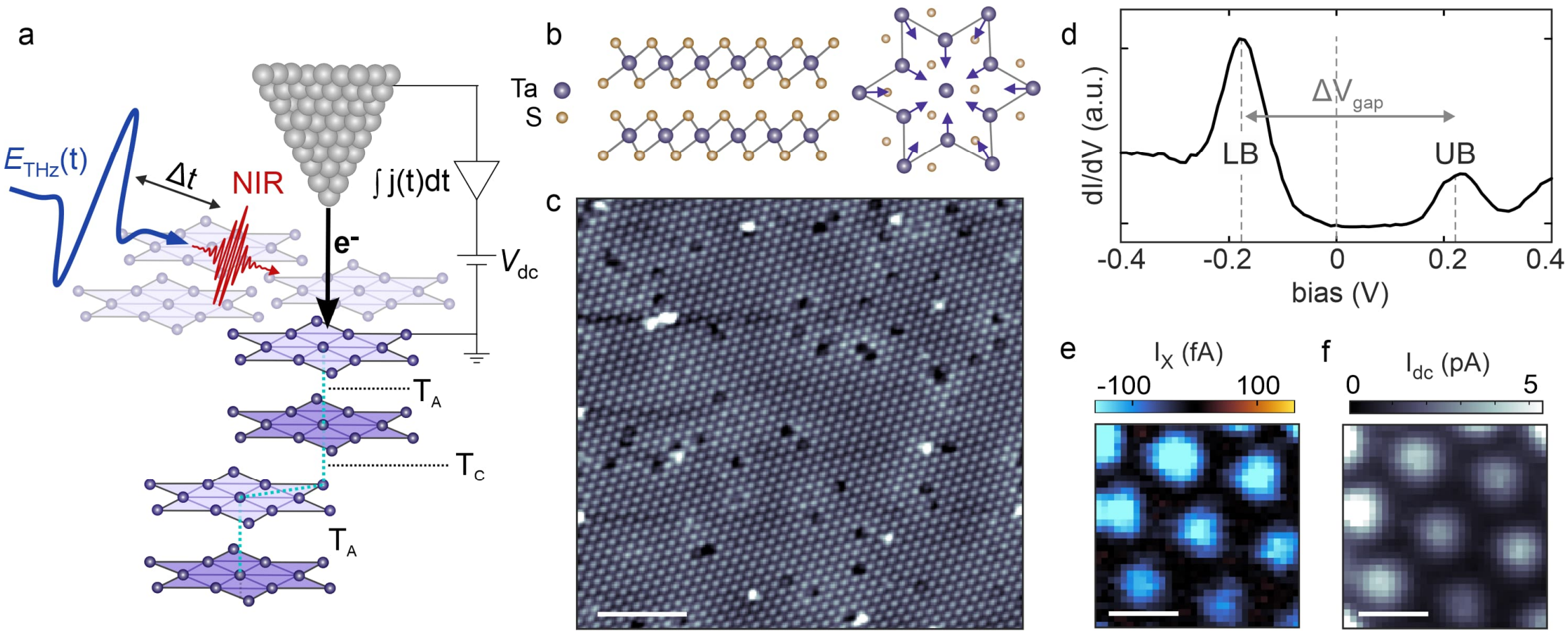

**Fig. 1: STM and THz-STM on 1T-$TaS_2$. a** Experimental scheme of THz-STM of photoexcited 1T-$TaS_2$. **b** Structure of 1T-$TaS_2$. **c** Constant-current STM image of the 1T-$TaS_2$ surface ($V_{dc}$ = −500 mV, $I_{SP}$ = 10 pA). Scale bar: 10 nm **d** Typical STS spectrum recorded over a defect-free area. **e** THz-STM image of the CDW and, **f** simultaneously recoded time-averaged current $I_{dc}$ at constant tip height ($V_{dc}$ = 50 mV, $E_{THz}$ = 780 V/cm). Scale bar: 1 nm

Figs. 1a and 1b show a sketch of the experimental setup and the structure of the CDW phase of 1T-$TaS_2$. Within each $TaS_2$ layer, 12 Ta atoms contract towards a 13$^{th}$ centre atom and form the characteristic SoD clusters. Fig. 1c shows a topographic STM image of a typical cleaved 1T-$TaS_2$ surface from our samples recorded without illumination at a temperature of 11 K. In the C-CDW phase, the SoD clusters arrange into a periodic $\sqrt{13} \times \sqrt{13}$ superlattice visible in the STM image.[28,34] We estimate that 2% of the SoD clusters contain a defect (irregular dark and bright spots in Fig. 1c). Scanning tunnelling spectroscopy (STS) recorded in defect-free regions reveals an insulating gap of $\Delta V_{gap} \approx 400$ meV (Fig. 1d), with the lower band (LB) and upper band (UB) of the CDW-split states typically located near $V_{LB} \cong -180$ mV and $V_{UB} \cong 220$ mV, respectively. This gap is characteristic of the dimerized $T_A$-terminated surface of a $T_AT_C$-stacked bulk (Fig. 1a), where interlayer dimerization gives rise to a more band-like insulating state.[34,35,37] We find similar STS spectra in various macroscopically different sample regions and for different samples cleaved from the same bulk crystal, indicating that $T_AT_C$-stacking with $T_A$-termination is the favored surface configuration in our samples.

To perform THz-STM measurements, we illuminate the STM junction with single-cycle THz-pulses (Fig. 1a) at a repetition rate of $f_{\mathrm{rep}} = 2$ MHz and with an incident peak electric field strength of $E_{\mathrm{THz}}$. The THz pulse train is chopped at a frequency $f_{\mathrm{chop}} = 607$ Hz for lock-in detection of the THz-induced currents (details are described in Methods). Figs. 1e and 1f show the spatial dependence of THz-induced change of the tunnelling current ($I_{\mathrm{X}}$) and the simultaneously recorded time-averaged current ($I_{\mathrm{dc}}$) during THz illumination. Here, $I_{\mathrm{X}}$ is the in-phase component of the lock-in signal as typically measured in lightwave-driven THz-STM.[13,14,26] Fig. 1e shows that the CDW superlattice can be resolved by THz-STM, with a spatial resolution similar to that obtained with $I_{\mathrm{dc}}$.

**Coherent oscillations of the local charge order**

Through lightwave-driven rectification, THz-STM can generate femtosecond tunnelling pulses that enable the probing of ultrafast dynamics. In Figs. 2 and 3, this is utilized to study the photoinduced dynamics of coherent collective modes at the 1T-$TaS_2$ surface. For this purpose, the STM junction is illuminated with 35-fs near-infrared (NIR) pump pulses polarized perpendicular to the STM tip. In this geometry, optical field enhancement of the NIR pulses is minimized to ensure a spatially homogeneous pump excitation. This allows spatial variations in the ultrafast response to be attributed to local variations of the sample properties rather than to a spatially varying excitation.

Figs. 2a and 2b show the THz-driven current $I_{\mathrm{X}}$ as a function of the THz-NIR time delay ($\Delta$t) for two macroscopically different sample areas in the range of few picoseconds after photoexcitation. Pronounced coherent oscillations of the THz-induced tunnelling current are observed. The corresponding Fast Fourier transform (FFT) spectra are shown in Figs. 2c and 2d. The tip position was fixed at a single SoD cluster during the measurements (pink stars in the STM images in Figs. 2e and 2f). The FFT spectra reveal a coherent mode at ~2.5 THz in both areas which matches the CDW amplitude mode (AM) in 1T-$TaS_2$, where the CDW amplitude oscillates synchronously with a zone-centre phonon corresponding to an atomic breathing motion of the SoD clusters.[3,30,31,48] Additionally, a distinct mode at ~1.3 THz and a peak below 1 THz are observed in area 2, which are absent in area 1.

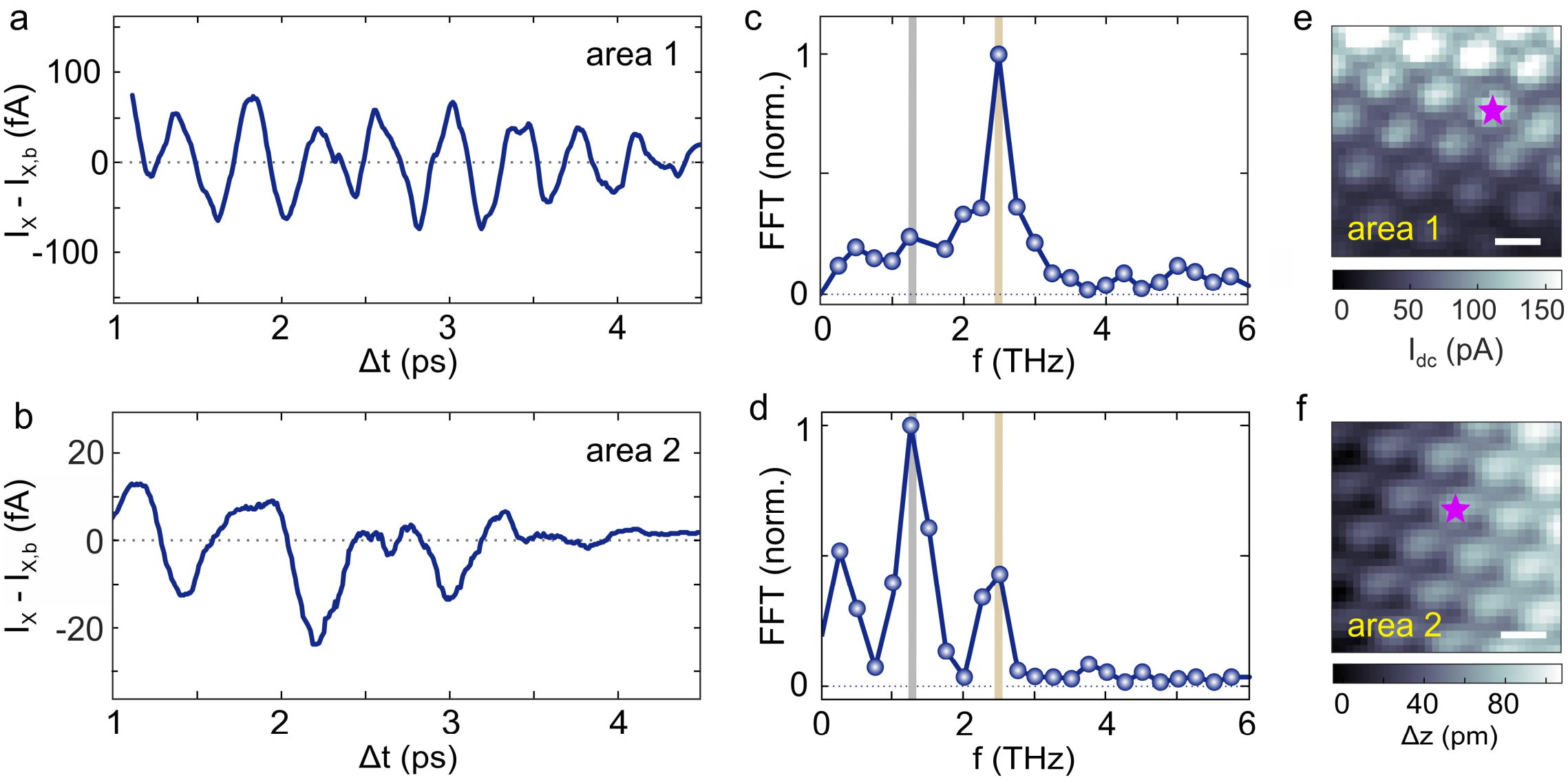

**Fig. 2: Local ultrafast dynamics in 1T-TaS$_2$. a,b** Coherent oscillations of the THz-induced current following NIR photoexcitation and **c,d** corresponding FFT spectra measured in two macroscopically different sample areas. **e** STM image of area 1 recorded at constant tip height set at $V_{\mathrm{dc}} = -200$ mV, $I_{\mathrm{SP}} = 10$ pA. **f** STM image of area 2 recorded at constant-current ($V_{\mathrm{dc}} = -500$ mV, $I_{\mathrm{dc}} = 10$ pA). Scale bars are 1 nm. Both images are recorded at negative delay $\Delta t = -100$ ps. The non-oscillatory background current $I_{\mathrm{X,b}}(\Delta t)$ is subtracted from the raw signal $I_{\mathrm{X}}$ (see Fig. S1 for raw data and baseline fits). STM set points are (a) $V_{\mathrm{dc}} = -100$ mV, $I_{\mathrm{dc}} = 10$ pA and (b) $V_{\mathrm{dc}} = 20$ mV, $I_{\mathrm{dc}} = 10$ pA. The NIR fluence and THz field are $\phi_{\mathrm{NIR}} = 29$ μJ/cm$^2$ and $E_{\mathrm{THz}} = 780$ V/cm, respectively.

The spatial resolution provided by THz-STM enables us to map such dynamics locally in the vicinity of local irregularities, as demonstrated in Fig. 3 for another sample region. Fig. 3a shows an overview of the CDW topography, whereas Fig. 3b shows a low-bias STM image recorded at $V_{\mathrm{dc}} = -10$ mV of the area marked by the white box in Fig. 3a. The CDW topography is locally enhanced with three adjacent SoD clusters appearing as bright protrusions. This suggests that the low-energy electronic structure close to the Fermi level $E_{\mathrm{F}}$ is locally perturbed at these superlattice sites. This is confirmed by the STS spectra in Fig. 3c, which show a modulation of the LDOS within the insulating gap at these positions (blue star, pink circle, and purple square). Specifically, the width and shape of the LB peak are altered, while its position remains unchanged. The UB peak broadens or splits at the purple-square position curve and is slightly shifted by about 30 mV at the pink-circle position. However, at the nearby SoD clusters (e.g. cyan cross and open diamond), the normal STS spectra that prevail in our samples are observed.

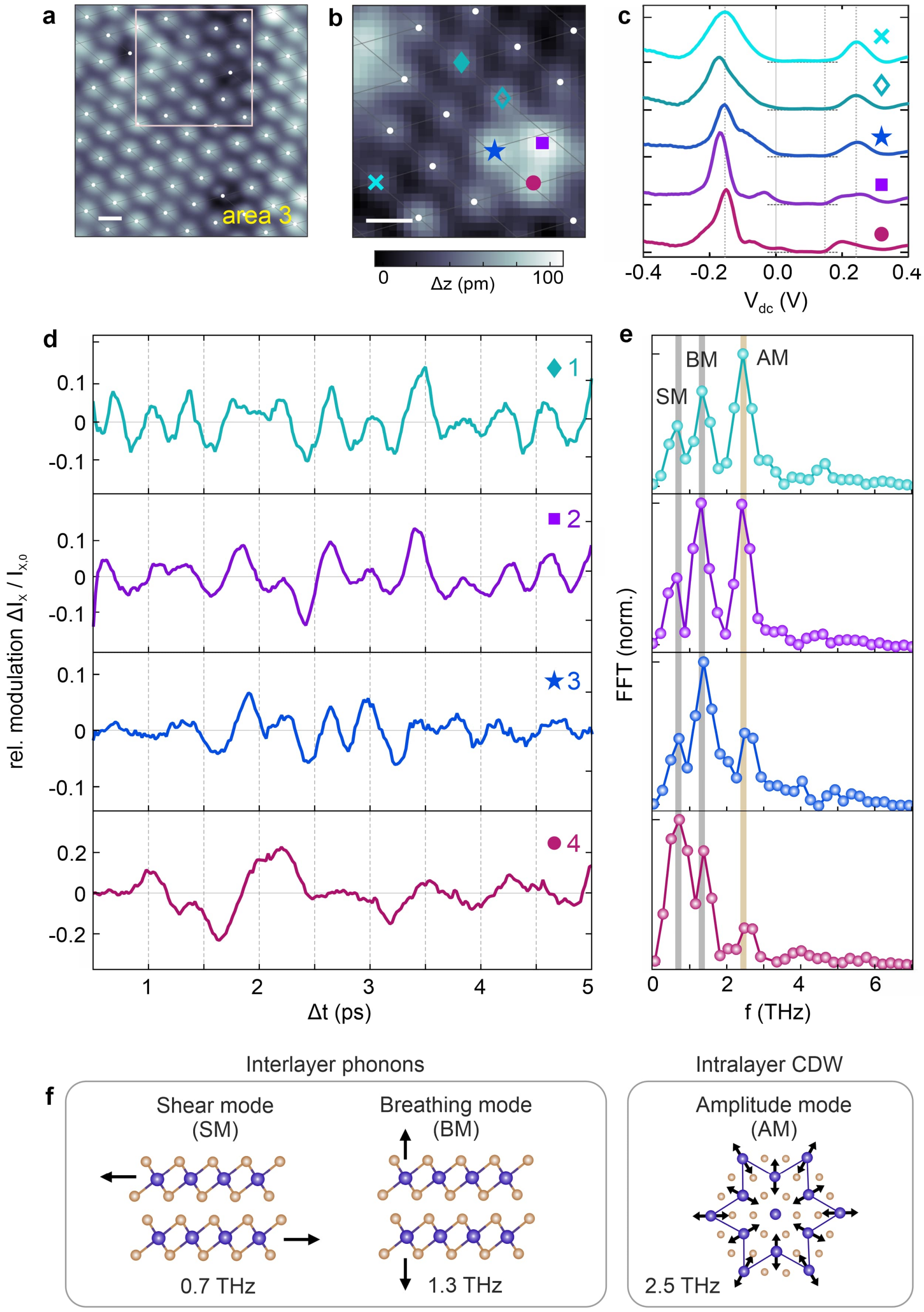

**Fig. 3: Spatial mapping of intra- and interlayer phonon dynamics**. **a** STM image of the sample region ($V_{dc}$ = −100 mV, $I_{dc}$ = 50 pA). **b** Low-bias STM image ($V_{dc}$ = −10 mV, $I_{dc}$ = 50 pA) of the area highlighted in (a). **c** STS spectra recorded in area 3. **d** Oscillatory part of the THz-induced tunnelling current recorded at the positions marked in 3b. The non-oscillatory baseline $I_{X,b}(\Delta t)$ is subtracted (see Fig. S1c) and the data is normalized to the baseline $I_{X,0}$ at long delays. **e** FFT spectra of the time-domain traces in (d). The AM at 2.5 THz and two low-frequency modes at 0.7 THz and 1.3 THz are resolved, with their amplitude ratio varying across the different SoD clusters. ($V_{dc}$ = −100 mV, $I_{dc}$ = 50 pA, $E_{THz}$ = 590 V/cm, $\phi_{NIR}$ = 29 μJ/cm$^2$, scale bars 1 nm). **f** Sketch of the observed phonon modes (phonon calculations are shown in section 2 in the SI).

Although it is difficult to unambiguously identify the nature of these LDOS modulations, a comparison with existing literature allows us to rule out common point defects, such as charged sulfur vacancies or titanium-tantalum substitution, which typically cause more localized and drastic changes to the low-energy LDOS.[38,40,47,49–52] Instead, it has been shown that the interlayer stacking order in 1T-$TaS_2$ can alter the low-energy LDOS of the insulating gap through stacking-dependent hybridization of the vertically aligned SoD orbitals.[33–37,53] The STS modulations observed in Fig. 3c are thus consistent with a localized stacking irregularity, and may reflect subtle modifications of the local orbital hybridization and interlayer hopping. At this stage, based on the present data, we simply identify the enhanced SoD clusters in Fig. 3c as a local perturbation of the low-energy electronic states at the surface.

To understand whether such local irregularity can influence the dynamics of collective modes, we locally probe the photoinduced response at different SoD clusters (positions marked in Fig. 3b). Fig. 3d shows the coherent oscillations measured in the THz-induced tunnelling current and Fig. 3e the corresponding FFT spectra. We observe again the AM peak at 2.5 THz, together with two low-frequency modes at 1.3 THz and ~0.7 THz. Interestingly, the ultrafast dynamics and the amplitude ratios of the modes vary across the different SoD clusters. While the AM is dominating in positions 1 (cyan diamond) and 2 (purple square), its amplitude decreases at position 3 (blue star) and almost vanishes at position 4 (pink circle). At the same time, the amplitude of the 0.7 THz mode is significantly enhanced.

To understand the nature of these modes, we calculate the bulk phonon modes in 1T-$TaS_2$ for different stacking configurations (see section 2 in the SI). For $T_A$-stacked $TaS_2$ in the CDW phase, where all layers are aligned in the $T_A$-sequence of the SoD clusters, the calculations yield two low-frequency phonon modes corresponding to an interlayer breathing mode (BM) at 1.24 THz and an interlayer shear mode (SM) at 0.73 THz as sketched in Figs. 3f and S3. These frequencies are in good agreement with the low-frequency modes measured in Fig. 3 and Fig. 2. For $T_AT_C$-stacked 1T-$TaS_2$ (Fig. S4), the reduced symmetry along the c-axis lifts the degeneracy of the SM, splitting it into two modes with frequencies of 0.68 THz and 0.75 THz corresponding to shear motion along different lateral directions. The SM frequency we measure in Fig. 3 is slightly below 0.7 THz. Increasing the frequency resolution in future measurements should enable the interlayer modes to be identified more precisely, providing a route to identifying the underlying interlayer stacking configuration.

The spatial variation in the amplitude ratios between the AM and interlayer SM and BM, as observed at nearby SoD clusters in Fig. 3, is not necessarily expected for spatially extended coherent lattice oscillations. Considering the rather homogeneous photoexcitation conditions, we attribute these variations to angstrom-scale changes of the surface condition (rather than to a locally varying pump). In particular, the significant increase of the SM amplitude observed at a single SoD cluster (pink circle, pos. 4) suggests that shear motion is locally enhanced at this CDW site, or that the LDOS at this site is more strongly modulated by the coherent SM. At the same time, the AM is strongly suppressed compared to the interlayer modes at this location. While it is difficult to quantify the mode-dependent detection mechanism at present (possible phonon detection mechanisms are discussed in section 7 of the SI), we suspect that local variations in interlayer coupling at each SoD cluster, caused by locally irregular stacking configurations, are the main cause of the varying phonon amplitudes. This is consistent with the in-gap LDOS variations observed in Fig. 3c, which we ascribed to variations in the interlayer orbital hybridization between nearby SoD clusters. Interestingly, as discussed in Fig. S8f of the SI and further below, area 3 is located close to a buried CDW domain boundary that is present below the surface. This corroborates the idea that the spatial inhomogeneities in the LDOS and phonon spectra in Fig. 3 originate from a spatially inhomogeneous stacking and interlayer coupling of SoD clusters.

**THz-induced metastable changes to the local CDW order**

The local detection of interlayer BM and SM phonons by ultrafast THz-STM motivates a closer examination of the role of interlayer hybridization and stacking order in our experiments. The weak interlayer bonding and nearly degenerate energies of different stacking configurations render the C-CDW phase of 1T-$TaS_2$ unusually susceptible to external perturbations. For example, excitation by voltage pulses in STM,[45,46] or single ultrafast laser pulses[3,11,42] has been shown to induce a metallic mosaic state associated with spatially varying stacking modifications on nanometre scales. This raises the question whether the ultrafast THz pulses used in our experiments act solely as an ultrafast probe, or whether they also modify the local charge order and interlayer stacking configuration. As presented below, several observations show that THz excitation of the tunnelling junction results in long-lived changes to the local electronic structure, which are attributed to local changes in the stacking and interlayer hybridization of SoD clusters. The lifetime of these changes range from quasi-stationary at strong THz excitation to milliseconds at weak THz fields comparable to those used during THz-STM operation.

To study the interaction of the tip-enhanced THz pulses with the C-CDW phase of 1T-$TaS_2$, we measure how the surface responds to excitation by the THz pulse train. We start with strong THz excitation with peak fields higher than those used during THz-STM operation. Figs. 4a and 4b show images of the 1T-$TaS_2$ surface before (top) and after (bottom) illumination with a train of strong THz pulses of ~1 s duration. The tip position was fixed (cyan star) and approximately $10^6$ THz pulses arrived at the sample. Under this condition, persistent THz-induced modifications of the CDW pattern are observed, including the local rearrangement of individual CDW domain boundaries (Fig. 4a) and the formation of locally disordered CDW regions (Fig. 4b). In Fig. 4a, two regions with distinct CDW phases are separated by a domain boundary (blue and white circles). Following THz excitation, a segment of the domain boundary is displaced, while several SoD clusters (pink circles with black arrows) rearrange to align with the CDW phase of the neighboring domain. In areas without a domain boundary but containing defects (Fig. 4b), the phase coherence of the CDW is locally disrupted after THz excitation, and small phase-shifted CDW domains embedded within an irregular domain-boundary network are created (yellow circles in bottom panel of Fig. 4b). As the second layer is unlikely to form identical domain boundaries or undergo identical SoD rearrangements and CDW phase shifts as the first layer, the observed changes in the CDW topography imply that the out-of-plane stacking of SoD clusters has been modified locally by the THz pulses. We note that at the used conditions, we do not observe the formation of a spatially extended metallic mosaic state. While further studies are necessary to fully establish the microscopic origin of the THz-induced CDW modifications, nonequilibrium hot carrier generation[44–46,54] and modified electronic screening[22,55] caused by the tip-enhanced THz fields may locally weaken the CDW and influence interlayer coupling, thereby facilitating lateral relative shifts of SoD clusters between adjacent layers. The observation that the changes predominantly occur near defects further suggests that they act as pinning centers and nucleation sites, decreasing the effective barrier for local CDW rearrangements.

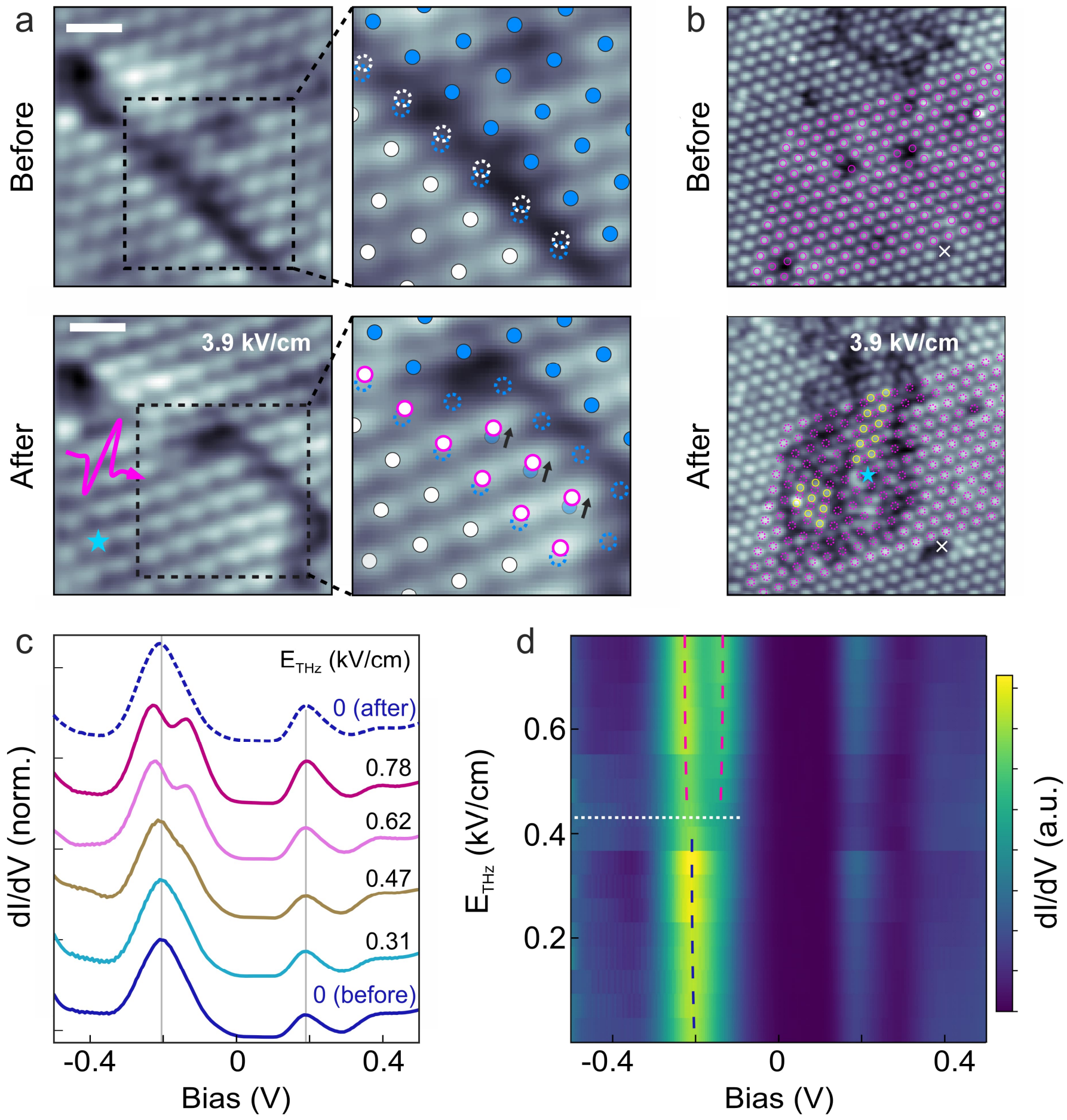

**Fig. 4: THz-induced modifications of LDOS and interlayer stacking**. **a,b** STM images recorded before (top) and after (bottom) THz illumination. **a** THz-induced motion of a CDW domain boundary. **b** Locally disordered CDW and loss of long-range phase coherence in the THz-excited area. Scale bars are (a) 2 nm and (b) 5 nm. STM images are recorded without THz illumination at $V_{dc} = -500$ mV. During THz illumination (duration ~1s, $E_{THz} = 3.9$ kV/cm), the tip position was fixed at the position of the cyan star at $V_{dc} = -10$ mV. Setpoint currents are (a) $I_{SP} = 60$ pA and (b) $I_{SP} = 50$ pA. The white cross marks the same defect as a guide to the eye. **c,d** STS spectra recorded during THz illumination for increasing THz field strength. The spectra are recorded in an area without a visible domain boundary (setpoint $V_{dc} = -500$ mV and $I_{SP} = 100$ pA). No persistent changes to the STS spectra or CDW topography are observed after the measurement.

At lower THz fields, we do not observe persistent changes to the CDW topography after THz excitation. Instead, fully reversible changes of the low-energy LDOS and the insulating gap can be observed by STS during THz illumination. Figs. 4c and 4d show dI/dV line spectra and a corresponding 2D colormap recorded at another sample location for THz field strength from 0 to 0.78 kV/cm. At low THz fields, the usual spectra of the $T_A$-terminated surface with peaks at $V_{\mathrm{LB}} \approx -200$ mV and $V_{\mathrm{UB}} \approx 200$ mV are observed. Above $E_{\mathrm{THz}} \gtrsim 0.4$ kV/cm, the LB splits into two peaks at $V_{\mathrm{LB},1} \approx -140$ mV and $V_{\mathrm{LB},2} \approx -230$ mV. The transition for observing the peak splitting exhibits a threshold-type behaviour around 0.4 kV/cm (white line in Fig. 4d). For THz fields above ~0.6 kV/cm, the STS spectra remain constant in the LB-split state. Interestingly, the observed LB changes are in very good agreement with the STS spectra previously observed for the $T_C$-terminated surface.[34,36] We thus attribute the observed STS changes to a THz-induced modification of the interlayer orbital hybridization between vertically aligned SoD clusters, giving rise to a quasi-stationary LDOS with features characteristic of stacking-dependent electronic configurations in 1T-TaS$_2$. Notably, the observed LB splitting is fully reversible, and the original T$_A$-stacking-type STS recovers within seconds after the THz pulse train is switched off (dashed blue STS in Fig. 4c).

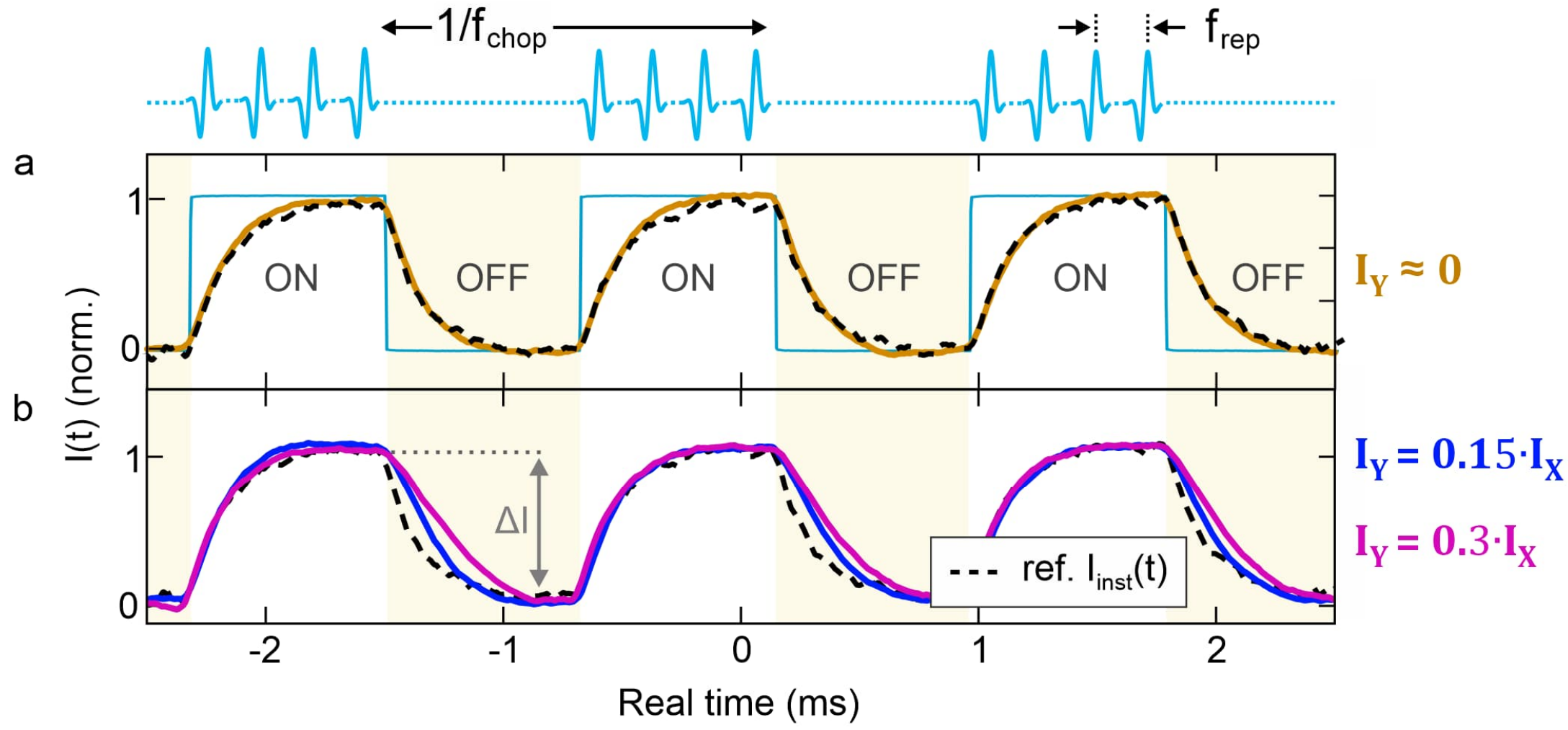

**Fig. 5: Real-time response of THz-induced tunnelling on 1T-TaS$_2$. a,b** Examples of real-time currents $I(t)$ measured on different sample locations and their respective quadrature contributions $I_Y$ of the lock-in signal. The black dashed lines show the instantaneous reference signal $I_{\mathrm{inst}}(t)$ as determined by the 1 kHz bandwidth of the STM preamplifier. The light blue line is the chopper signal at the lock-in reference input. The signal $I_Y$ is a hallmark of a non-instantaneous response and evidence for millisecond dynamics. Data recorded at $E_{\mathrm{THz}} = 390$ V/cm and $I_{\mathrm{SP}} = 50$ pA, $V_{\mathrm{dc}} = -50$ mV (a), $V_{\mathrm{dc}} = -50$ mV, $I_{\mathrm{SP}} = 30$ pA (b, blue) and $V_{\mathrm{dc}} = -10$ mV, $I_{\mathrm{SP}} = 50$ pA (b, pink). The data in (b) is measured at the positions marked in Figs. 6c (blue) and S5 (pink), respectively.

**Real-time dynamics of THz-induced metastability at weak excitation**

The LDOS changes observed during weak THz illumination shown in Figs. 4c and 4d indicate that a nonequilibrium state is created whose lifetime must exceed the 0.5 μs inverse repetition rate of the THz pulse train. Conversely, the complete recovery of the LDOS immediately after blocking the THz pulse train indicates a lifetime below a few seconds. Importantly, the THz field strengths at which such reversible LDOS changes are observed are comparable to those used during ultrafast THz-STM measurements.

To study the dynamics of these THz-induced LDOS changes and to understand their role in the ultrafast THz-STM measurements, we measure the millisecond real-time response of the tunnelling current to the chopped THz pulse train with an oscilloscope. Figs. 5a,b show representative real-time traces of the tunnelling current $I(t)$ recorded in different sample regions. The light blue square wave in Fig. 5a shows the modulation of the THz beam by the chopper with a period $1/f_{\text{chop}} \approx 1.6$ ms. Note that the timescale on which a time-delayed tunnelling response can be measured is limited by the bandwidth of the STM preamplifier (1 kHz in the present experiments; see detailed discussion in section 5 in the SI). The black dashed lines show the resulting instrument response function $I_{\text{inst}}(t)$ of the STM electronics, which is obtained by measuring the THz-induced modulation of ultrafast photoemission from the tip (details see Methods and SI).

While the tunnelling current in Fig. 5a (orange) directly follows the instrument response $I_{\text{inst}}(t)$, the curves in Fig. 5b (blue and pink) exhibit a delayed decay when the THz beam is blocked after the ON→OFF transition. Such a delayed response is characteristic for a metastable excitation of the sample with a lifetime exceeding the preamplifier bandwidth, similar to recent THz-STM measurements on $WTe_2$.[26] In some cases, we find even stronger temporal distortions of $I(t)$, where a delayed response also occurs at the OFF→ON transition. In this case, the accumulation of multiple THz pulses at the laser repetition rate generates a long-lived excitation in the sample which does not reach a steady-state during the 0.8 ms chopper ON period.

Importantly, the delayed millisecond response of the tunnelling conductance after blocking the THz pulse train observed in Fig. 5b is inconsistent with the picture for ultrafast THz-lightwave-driven rectification. Instead, it can be explained by reversible, quasi-stationary changes to the LDOS induced by the THz pulses, as observed in Figs. 4c,d. In this case, the THz-induced current modulation $\Delta I$ (see Fig. 5) can be attributed to a different LDOS in the 'ON' and 'OFF' states at the measurement bias, consistent with the creation of a metastable state, rather than to ultrafast lightwave-driven currents.

In general, the THz pulses can drive ultrafast lightwave-rectification and excite quasi-stationary changes of the LDOS simultaneously during a single measurement. We thus write $\Delta I = \Delta I_{\mathrm{LW}} + \Delta I_{\mathrm{LDOS}}$, where $\Delta I_{\mathrm{LW}}$ is the rectified current measured in conventional lightwave-driven THz-STM and $\Delta I_{\mathrm{LDOS}} = I_{\mathrm{THz,on}} - I_{\mathrm{THz,off}}$ is the current arising from THz-induced changes of the LDOS. Importantly, the two contributions, $\Delta I_{\mathrm{LW}}$ and $\Delta I_{\mathrm{LDOS}}$, originate from THz-induced tunnelling currents that occur on drastically different timescales. Whereas $\Delta I_{\mathrm{LW}}$ consists of femtosecond current pulses that allow to probe ultrafast dynamics, $\Delta I_{\mathrm{LDOS}}$ reflects a THz-induced transition to a long-lived metastable state. Therefore, in addition to probing ultrafast dynamics, THz-STM allows to directly image the formation of metastable states.[26]

**Imaging local metastability in 1T-TaS$_2$ by THz-STM**

To explain how THz-STM can be used to image metastability on millisecond time and angstrom length scales, we discuss the lock-in signals measured in THz-STM. According to the principles of lock-in detection, a delayed time-domain response of $I(t)$ as shown in Fig. 5b leads to a non-zero quadrature component $I_{\mathrm{Y}} \neq 0$ (see Methods and Figs. S6 for a detailed explanation).[26] We calibrate the lock-in phase so that $I_{\mathrm{Y}} = 0$ for the instantaneous reference signal $I_{\mathrm{inst}}\,(t)$ (dashed black line in Fig. 5). For pure THz-lightwave-driven rectification, $\Delta I = \Delta I_{\mathrm{LW}}$ and the tunnelling conductance will follow $I_{\mathrm{inst}}\,(t)$, so that $I_{\mathrm{Y}} = 0$. However, if the real-time waveform of $I(t)$ changes due to millisecond-scale dynamics, as shown in Fig. 5b, the quadrature signal will usually be non-zero ($I_{\mathrm{Y}} \neq 0$). Importantly, while the in-phase signal $I_{\mathrm{X}}$ can contain contributions from both $\Delta I_{\mathrm{LDOS}}$ and $\Delta I_{\mathrm{LW}}$, $I_{\mathrm{Y}}$ only contains contributions from $\Delta I_{\mathrm{LDOS}}$ due to slow millisecond LDOS modulations.[26] Therefore, the observation of a non-zero quadrature signal $I_{\mathrm{Y}} \neq 0$ during THz-STM operation provides direct evidence for the formation of a THz-induced metastable state with an altered LDOS.

In Fig. 6, we use this to study the spatial variation of long-lived millisecond-scale THz-induced LDOS changes in 1T-TaS$_2$. For this purpose, we image the quadrature signal $I_{\mathrm{Y}}$ together with $I_X$ during THz-STM operation. The $I_Y$ images yield differential maps of the THz-induced LDOS changes, i.e., they reveal sample regions in which the metastable LDOS changes are particularly pronounced (at the used $V_{dc}$). Interestingly, we find that a non-zero current $I_Y$ predominantly occurs near defects (white circles in the topographic STM images). However, the THz-induced LDOS changes are not localized directly at defects, but extend over spatially inhomogeneous nanoscale regions in their close vicinity. While $I_Y$ unambiguously results from a millisecond-delayed tunnelling response, we note that also $I_X$ can be dominated by $\Delta I_{\mathrm{LDOS}}$, meaning the image contrast in the $I_{\mathrm{X}}$ maps can also be dominated by the metastable response.

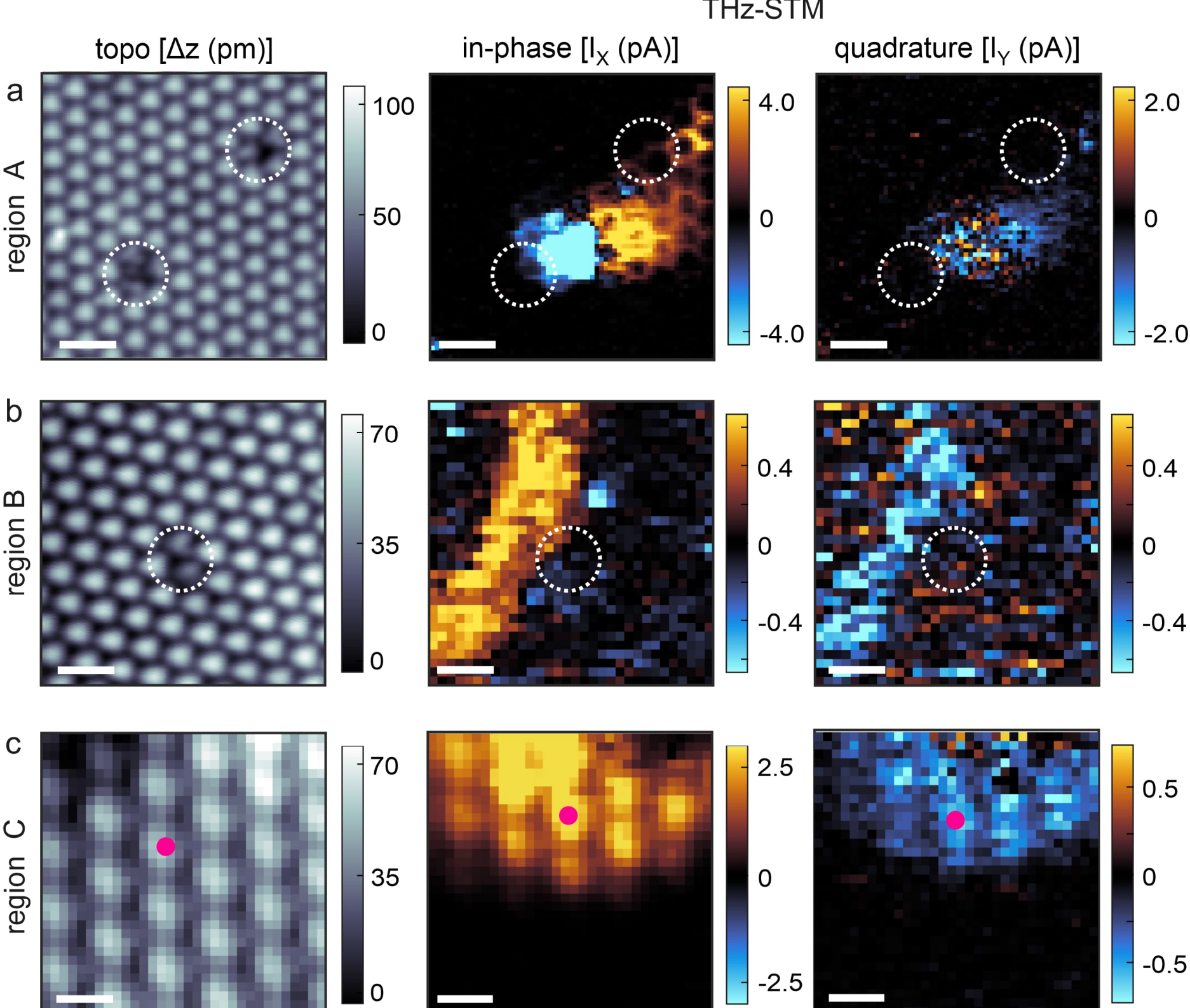

**Fig. 6: Lock-in mapping of THz-induced metastability**. **a-c** Topographic (left) and THz-STM images of the 1T-$TaS_2$ surface. Both the in-phase ($I_X$, middle) and quadrature components ($I_Y$, right) of the lock-in signal are plotted. The signal $I_Y$ allows to identify sample regions that exhibit a THz-induced tunnelling response on the millisecond timescale. These are typically found near defects (white circles), but are not directly localized at the defects. THz-STM images are taken at $E_{THz}$ = 390 V/cm and (a) $I_{SP}$ = 60 pA, $V_{dc}$ = −50 mV, (b) $I_{SP}$ = 90 pA, $V_{dc}$ = −100 mV, and (c) $I_{SP}$ = 40 pA, $V_{dc}$ = −100 mV. Topographic STM images are taken at $V_{dc}$ = −500 mV (a,b) and $V_{dc}$ = −100 mV (c). Scale bars are 2 nm (a,b) and 1 nm (c). The pink circle in (c) shows the position where the blue real-time trace in Fig. 5b was recorded.

Notably, the differential THz-STM images can reveal "hidden" domain boundaries (Fig. 6b) or nanoscale domains (Fig. 6a,c) that are not visible in the static CDW topography. Note that the observed images cannot be explained by THz-induced charging of local defects or THz-induced band bending, which would be more localized around the defects. Moreover, at realistic junction capacitances and tunnelling resistances, parasitic capacitive coupling cannot account for the strongly localized millisecond-scale tunnelling response. Instead, together with the results in Fig. 4, the THz-STM images in Fig. 6 strongly suggest that THz-induced local stacking modifications, with lifetimes in the millisecond range, play a key role in the formation of THz-induced metastable states observed by THz-STM.

**DISCUSSION**

The above results highlight the dual role of THz excitation in our experiments. On the one hand, it excites localized metastable states of altered interlayer hybridization at the 1T-$TaS_2$ surface. On the other hand, it serves as a femtosecond bias to probe ultrafast NIR-photoinduced dynamics. As we show in section 6 of the SI for the measurements in area 3 (see Fig. 3), the THz-driven current $I_{\mathrm{X}}$ consists not only of ultrafast rectification ($\Delta I_{\mathrm{LW}}$, THz as "probe"), which enables to probe photoinduced dynamics, but also of contributions from long-lived LDOS changes ($\Delta I_{\mathrm{LDOS}}$, THz as "pump"), revealing the presence of a THz-induced metastable state during ultrafast THz-STM measurements. This is evident from several observations, including a non-zero quadrature signal $I_{\mathrm{Y}}$ (Fig. S8), the scaling of $I_{\mathrm{X}}$ and $I_{\mathrm{dc}}$ with $E_{\mathrm{THz}}$, as well as the unconventional polarity behaviour of $I_{\mathrm{X}}$ (Fig. S9a-d), all of which are highly spatially inhomogeneous. In particular, the characteristic dependence of $I_{\mathrm{X}}$ on $E_{\mathrm{THz}}$ and $V_{\mathrm{dc}}$ can be reproduced by numerical simulations of the THz-induced current $\Delta I = I_{\mathrm{LW}} + \Delta I_{\mathrm{LDOS}}$, where contributions from both $\Delta I_{\mathrm{LW}}$ and $\Delta I_{\mathrm{LDOS}}$ are necessary to fully reproduce the experimental data (see Fig. S9c-f).

Moreover, our results strongly suggest that the metastable state includes THz-induced changes to the interlayer hybridization and stacking of SoD clusters. In area 3, this is supported by the low-bias STM image shown in Fig. S8f, which reveals that the probed region is located next to a buried CDW domain boundary that is present below the surface (Fig. S8f). This is consistent with our previous observations (Fig. 4 and 6) that THz-induced metastability is predominantly observed near defects and domain boundaries. We thus conclude that the coherent phonon modes in Fig. 3 are not governed by the equilibrium ground state of the 1T-$TaS_2$ surface, but are photoexcited within a THz-induced metastable state of locally altered interlayer coupling.

A key finding of our work is that ultrafast dynamics and metastability can be probed simultaneously, enabling a new route for ultrafast spectroscopy of metastable states. We summarize our results in Fig. 7 and the Supplementary Figure S10, where we qualitatively describe the system's dynamics by introducing a two-dimensional potential energy surface (PES) defined by the intralayer CDW order parameter $\Psi_{\mathrm{CDW}}$ and an interlayer stacking order parameter $\Psi_{\mathrm{IL}}$. Here, $\Psi_{\mathrm{CDW}}$ represents the CDW amplitude of the symmetry-broken ground state of 1T-$TaS_2$ at low temperature, and $\Psi_{\mathrm{IL}}$ represents the interlayer coupling and hybridization of SoD orbitals. While $\Psi_{\mathrm{CDW}}$ is assumed to be independent on the incident THz field, $\Psi_{\mathrm{IL}}$ can be modified by THz excitation and thus depends on $E_{\mathrm{THz}}$. As $E_{\mathrm{THz}}$ increases, the intralayer CDW order remains in the symmetry-broken C-CDW phase, but the system evolves along $\Psi_{\mathrm{IL}}$ and gains energy due to THz excitation and modified interlayer coupling, until it reaches a metastable configuration (process 1, cyan arrow in left projection). Such a transition is consistent with the observation in Fig. 4d that a threshold THz field is required to induce long-lived changes to the insulating gap in the LDOS.

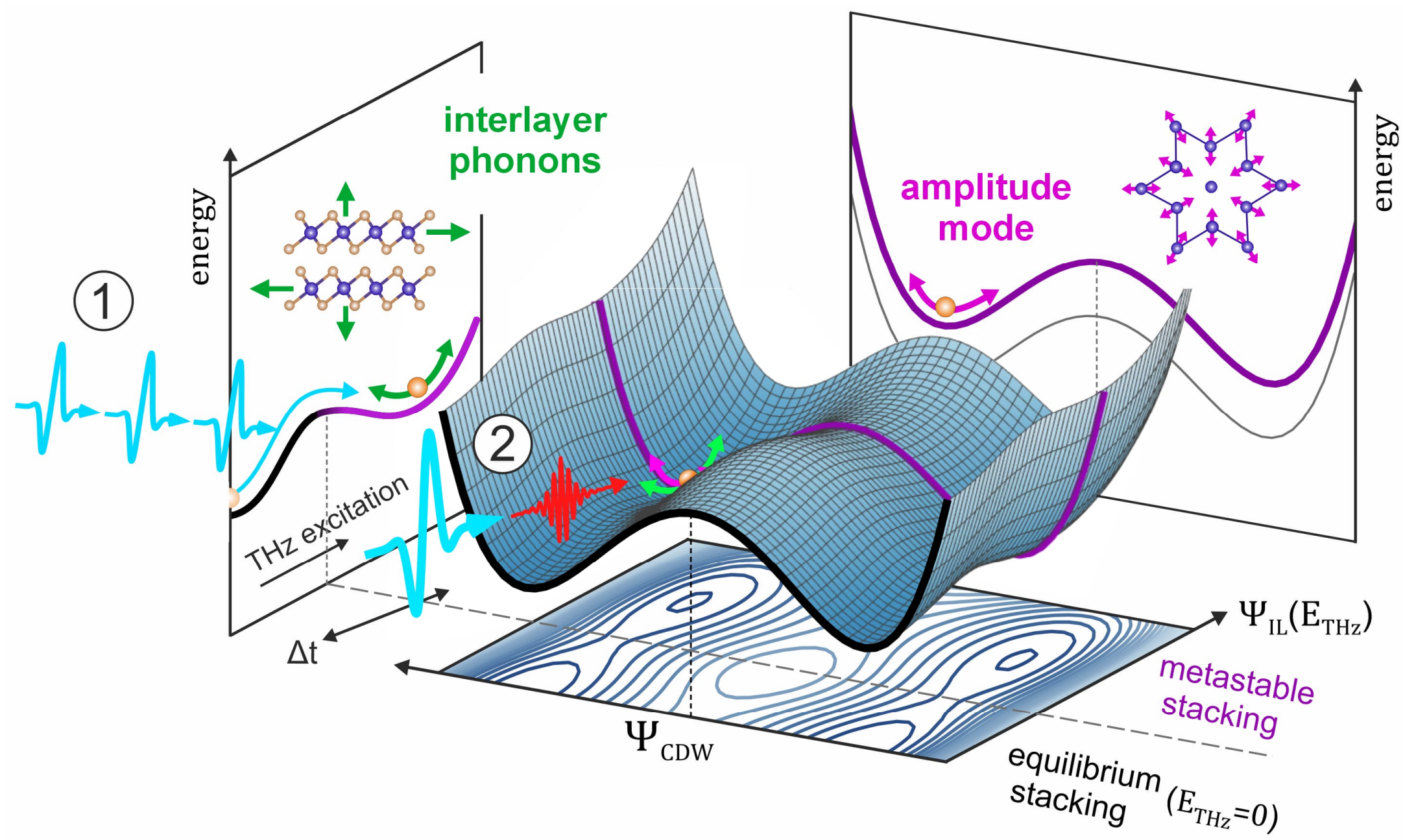

**Fig. 7: Ultrafast dynamics in a metastable energy landscape.** Illustration of the conceptual two-dimensional potential energy surface (PES) spanned by the intralayer CDW order parameter $\Psi_{CDW}$ and the interlayer parameter $\Psi_{IL}$. Along $\Psi_{CDW}$, the PES exhibits the double-well potential of the symmetry-broken C-CDW phase (thick black curve at $E_{THz} = 0$). Along $\Psi_{IL}$, THz excitation modifies the interlayer potential and can drive the system into a localized metastable stacking configuration (step 1, left projection), in which the symmetry-broken CDW potential is preserved (thick purple curve). NIR photoexcitation can then induce phonon oscillations within the metastable state, which are probed by ultrafast THz-lightwave-driven rectification (step 2). While the AM oscillates along $\Psi_{CDW}$ (back projection), the interlayer modes (ILM) coherently modulate $\Psi_{IL}$ (left projection). We emphasize that this PES is a conceptual summary of the results. In particular, it likely does not represent the global energy landscape of the system, but may vary between individual CDW superlattice sites. The dual role of the THz pulses in driving metastability (step 1) and probing ultrafast dynamics (step 2) is illustrated further in Fig. S10 in the Supplementary Material.

The THz fields used for ultrafast lightwave-driven STM are typically above this threshold field, which is consistent with the observation of a metastable millisecond-scale response during ultrafast THz-STM measurements. We thus conclude that ultrafast NIR-photoexcitation induces coherent oscillations within this THz-induced metastable state, which are probed by THz-lightwave-driven rectification (process 2). Intriguingly, as explained in Fig. S10, process 1 and 2 can happen sequentially within the THz pulse train. While the AM, which describes a coherent oscillation along $\Psi_{CDW}$ (back projection in Fig. 7), should be independent of $\Psi_{IL}$, the interlayer shear and breathing modes at 0.7 THz and 1.3 THz will depend on $\Psi_{IL}$, as confirmed by the stacking-order dependent phonon calculations (Figs. S2-4 and table T1 in the SI). Hence, we expect these modes to differ between the metastable and ground state of the 1T-$TaS_2$ surface. Increasing the frequency resolution of our measurement in future work should enable us to identify different mode frequencies more precisely. This would provide a spectroscopic approach to probe the local interlayer stacking via specific phonon fingerprints, and for studying the interplay of stacking order and metastability in 1T-$TaS_2$ and other layered quantum materials with angstrom resolution.

Finally, we emphasize that in the presence of stacking irregularities, domain boundaries and defects, the creation of a metastable stacking configuration does not require a global displacement of the entire 1T-$TaS_2$ surface layer. Instead, the THz-induced changes to the interlayer CDW order and SoD coupling observed in this study can be highly localized and likely involve more subtle interlayer shifts, which may vary between individual SoD clusters. The PES in Fig. 7 thus does not necessarily represent the global CDW order, but rather reflects a local energy landscape defined by the local environment near a single SoD cluster.

More broadly, our work establishes THz-STM as a powerful platform for imaging and manipulating nonequilibrium quantum phases at their intrinsic spatiotemporal scales, ranging from ultrafast to metastable timescales. The ability of THz-STM to simultaneously access local defects, nanoscale disorder, ultrafast dynamics, and metastability promises to offer new insights into the multi-scale mechanisms underlying metastable phase formation and characterization. We anticipate that our findings will stimulate further experimental and theoretical efforts to understand the complex interplay between local structure, collective excitations, and emergent nonequilibrium behaviour in quantum materials.

## METHODS

### Sample and tip preparation

Samples were cleaved with scotch tape from a commercial 1T-$TaS_2$ bulk crystal (HQ graphene) under high vacuum conditions ($1 \cdot 10^{-8}$ mbar base pressure) at room temperature. The freshly cleaved sample was introduced into the STM chamber immediately after exfoliation to minimize surface contamination. An electrochemically etched tungsten tip was used for all measurements. Tip conditioning was performed via controlled indentations and voltage pulses on a clean Au(111) surface. The tip condition was checked by imaging and tunnelling spectroscopy on Au(111).

### Optical setup

The laser setup is based on a commercial high-power femtosecond laser (Light Conversion Carbide, 80 W) delivering 1030 nm near-infrared (NIR) pulses with 200 fs duration at a repetition rate of $f_{\mathrm{rep}} = 2$ MHz. The NIR pulses are spectrally broadened and further compressed to ~35 fs using a multiplate compressor. The compressed pulses are split into two beam paths at 95:5 intensity ratio. The low-power NIR beam is used for optical excitation of the sample in the pump-probe measurements. The high-power path is used to generate linearly polarized single-cycle THz pulses from a rotating spintronic THz emitter (STE, TeraSpinTec GmbH).[56] Residual NIR light is filtered by a combination of indium-tin-oxide (ITO)-coated glass and a Germanium wafer before the THz pulses are coupled into the STM. The low-power NIR beam is collinearly overlapped with the THz pulses at a second ITO glass and the NIR-THz delay is controlled by a motorized translation stage. To enable lock-in detection of the THz-induced current, the high-power NIR beam used for THz generation is modulated by a mechanical chopper at a frequency of $f_{\mathrm{chop}} = 607$ Hz. The waveform and amplitude of the incident THz electric field are measured by electro-optic sampling in a reference beam path outside the STM. The tip-enhanced THz waveform is measured inside the STM via photoemission sampling from the free-standing tip (~1 mm retraction from the sample) [16,57]. In both cases, 35 fs NIR pulses are used as sampling pulses. Time zero is defined by the zero-crossing of the measured THz near-field waveform (see Fig. S12).

**STM setup**

Measurements were performed in a customized commercial low-temperature STM (Unisoku USM-1400) under ultra-high vacuum (UHV) conditions ($\sim 5 \cdot 10^{-11}$ mbar base pressure) at 11 K. The STM bias is applied to the sample and the current is measured via the grounded tip using a current preamplifier (Femto-DLPCA) at $10^9$ V/A gain and 1 kHz bandwidth. STS spectra are acquired in constant height mode with a bias modulation amplitude of 20 mV at a frequency of 781 Hz. The junction is illuminated by the collinearly propagating NIR and THz pulses via a motorized off-axis parabolic mirror (OAPM, Au-coated, focal length 35 mm) mounted inside the cold STM head. The OAPM can be moved in all three dimensions for precise focusing on the STM tip. The THz pulses are polarized parallel to tip axis. For NIR-pump - THz-probe STM measurements, the NIR pulses are polarized parallel to the sample surface to minimize optical field enhancement and achieve spatially homogeneous photoexcitation of the sample. For photoemission sampling, the NIR pulses are polarized along the tip axis. The THz-induced tunnelling current is measured with an external lock-in amplifier (Stanford Research Systems, SRS830) referenced to $f_{\mathrm{chop}}$.

**Lock-in phase calibration**

Before each measurement series, the phase of the external lock-in amplifier was calibrated using a reference signal that responds instantaneously to the chopped THz pulse train. This signal was obtained by monitoring the THz-induced modulation of the NIR-driven photoemission current from the free-standing STM tip. The lock-in reference phase was then adjusted such that a THz-induced increase of the photoemission current - corresponding to THz-induced electron transport from tip to sample - yields a positive in-phase signal $I_{\mathrm{X}}$, with the quadrature signal $I_Y \approx 0$. The corresponding time-dependent current $I(t)$ is measured at the output of the current preamplifier with the oscilloscope and provides the reference signal in Fig 5.

**Statistical Analysis and Data Analysis**

This study presents local measurements obtained using ultrafast scanning tunnelling microscopy under cryogenic conditions. The number of measurements that can be taken at a particular defect site or sample area is often limited by the two-day hold time of the STM bath cryostat. An integration time of $t_{\mathrm{int}} = 100$ ms was used for the lock-in measurements of $I_X$ and $I_Y$. Depending on the signal-to-noise ratio, each measurement is averaged over several scans for a set of experimental parameters and at a given tip location. The measurements were repeated several times over several months on different samples and at various locations. A metastable response is observed frequently for different locations on all investigated samples. The CDW amplitude mode and interlayer phonon modes have been observed in more than 10 independent measurements at different sample locations. The simulations and data analysis was carried out using Python, MATLAB and WSxM.[58] All pump-probe traces shown in the manuscript have been processed by fitting and subtracting their non-oscillatory background. For illustrative purposes, the time-domain data has then been smoothed. The FFT spectra were calculated from the unsmoothed time-domain traces using a Hamming window to ensure the reliability of the spectral analysis.

**AUTHOR INFORMATION**

**Corresponding Author**

*L.E.P.L.: lopez@fhi-berlin.mpg.de

*M.M.: m.mueller@fhi-berlin.mpg.de

**Present Addresses**

# F.S.: CIC nanoGUNE, Tolosa Hiribidea 76, 20018 San Sebastian, Spain.

† M.M.: Institute of Physics, University of Bonn, 53115 Bonn, Germany

**ACKNOWLEDGEMENTS**

The authors thank L. Rettig, C. Nicholson, A. Paarmann, H. Wiedenhaupt and D.B. Shin for fruitful discussions and H. Kirsch for tip preparation support. We gratefully acknowledge financial support from the Max Planck Society. M.M. is grateful for funding through the ERC H2024 StG project FASTOMIC/Grant No. 101165707. J. Q. and A. R. acknowledge support from the Cluster of Excellence "CUI: Advanced Imaging of Matter" of the Deutsche Forschungsgemeinschaft (DFG) - EXC 2056 - Project ID No. 390715994, the Max Planck-New York City Center for Non-Equilibrium Quantum Phenomena, as well as from the European Union under the ERC Synergy Grant UnMySt (HEU GA No. 101167294). The Flatiron Institute is a division of the Simons Foundation.

**CONFLICT OF INTEREST**

The authors declare no conflict of interest.

# Supplementary Information

# Ultrafast dynamics of local charge order in a THz-induced metastable quantum state

Luis E. Parra López,[1*] Jingkai Quan,[2] Alkisti Vaitsi,[1] Vivien Sleziona,[1] Fabian Schulz,[1#] Angel Rubio,[2,3] Martin Wolf,[1] and Melanie Müller[1†*]

*[1]Department of Physical Chemistry, Fritz Haber Institute of the Max Planck Society, 14195 Berlin, Germany*
*[2]Max Planck Institute for the Structure and Dynamics of Matter, 22761 Hamburg, Germany*
*[3]Initiative for Computational Catalysis, Flatiron Institute, Simons Foundation, New York City, NY 10010, USA*

** Corresponding authors: lopez@fhi-berlin.mpg.de, m.mueller@fhi-berlin.mpg.de*

*Present Addresses:*
*# F.S.: CIC nanoGUNE, Tolosa Hiribidea 76, 20018 San Sebastian, Spain*
*† M.M.: Institute of Physics, University of Bonn, 53115 Bonn, Germany*

## 1. Pump-probe raw data

Fig. S1 shows example raw data of pump-probe traces for the areas 1 and 2 (Fig. 2) and area 3 (Fig. 3). For all pump-probe traces shown in the manuscript, the non-oscillatory background is fitted and subtracted (orange lines in Fig. S1). The time-domain data is slightly smoothed for displaying purposes only (see Fig. S1). The Fast Fourier spectra are obtained by applying the Fourier transform to the background-corrected, unsmoothed data.

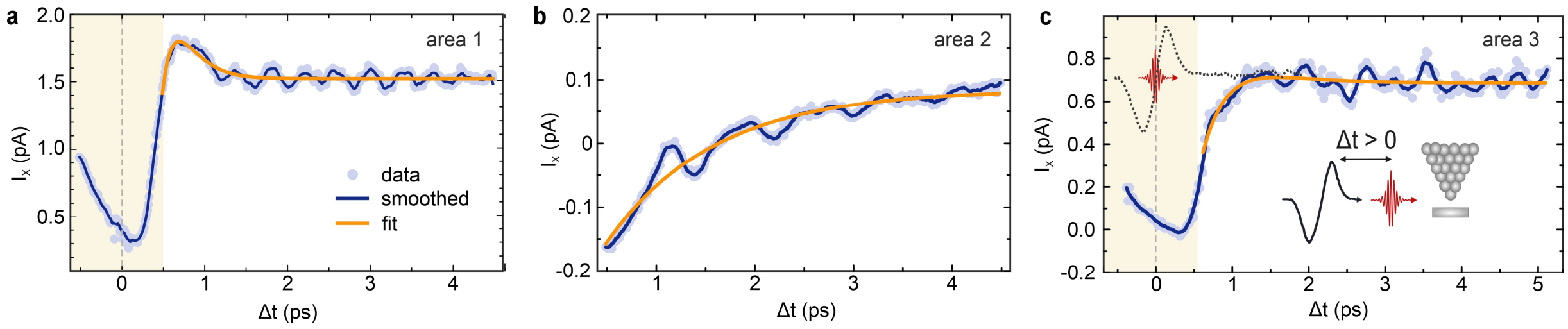

**Fig. S1: Coherent phonon dynamics in 1T-TaS₂. a-c** THz-NIR cross-correlation measurements recorded on area 1 (a), area 2 (b) and area 3 (c) from the main manuscript, showing coherent oscillations of the tunnelling current shortly after photoexcitation. The data is presented raw (blue disks) and smoothed (blue line) for each curve. The non-oscillatory baseline is fitted by a double exponential function (orange line). The traces were recorded at an STM setpoint of (a) $V_{\mathrm{dc}} = -100$ mV, $I_{\mathrm{dc}} = 10$ pA, $E_{\mathrm{THz}} = 780$ V/cm, (b) $V_{\mathrm{dc}} = 20$ mV, $I_{\mathrm{dc}} = 10$ pA, $E_{\mathrm{THz}} = 780$ V/cm and (c) $V_{\mathrm{dc}} = -100$ mV, $I_{\mathrm{dc}} = 50$ pA, $E_{\mathrm{THz}} = 590$ V/cm and with $\phi_{NIR} = 29$ μJ/cm$^2$.

## 2. Calculation of phonon modes

To investigate the vibrational modes of 1T-$TaS_2$, we performed density functional theory (DFT) calculations using the all-electron numeric atom-centered orbital code *FHI-aims*.[1,2] The self-consistent calculations are performed using the '*light*' basis set, the PBE[3] exchange-correlation functional, and the nonlocal many-body dispersion (MBD-NL) approach[4] to account for van der Waals interaction. Phonon calculations were performed with *FHI-vibes*[5] interfaced with *PHONOPY*.[6,7] For perfect $T_A$-stacking 1T-$TaS_2$, we use a $6 \times 6 \times 2$ supercell and a $3 \times 3 \times 6$ k-grid in phonon calculation. For $T_A$-stacking $TaS_2$ in the charge density wave (CDW) phase, whose unit cell enlarged to $\sqrt{13} \times \sqrt{13} \times 1$ and contains 39 atoms, we used a $2 \times 2 \times 2$ supercell and a $2 \times 2 \times 2$ k-grid in phonon calculation. For $T_AT_C$-stacking CDW $TaS_2$, whose unit cell contains four layers of $TaS_2$, we used a $1 \times 1 \times 1$ cell and a $3 \times 3 \times 2$ k-grid in phonon calculation.

The resulting phonon dispersions are shown in Fig. S2. We focus on the low-frequency range, which is relevant to the experiments. As illustrated in Fig. S2a, the perfect $T_A$-stacking 1T-$TaS_2$ exhibits an imaginary mode between $\Gamma - \mathrm{M}$, indicating the CDW instability. Figs. S2b and S2c show the phonon dispersions of the CDW phase of $TaS_2$ with different stacking orders. In these cases, the phonon bands are folded together due to the increase of the supercell size. The frequencies of the interlayer vibration modes for both $T_A$ and $T_AT_C$ stacking of CDW $TaS_2$ are listed in Tab. S1. The corresponding displacement patterns are shown in Figs. S3 and S4. The frequency of the interlayer breathing mode of the $T_AT_C$ stacking structure is 1.27 THz, in good agreement with the ~1.3 THz peak observed in the FFT spectra of the THz-induced current (Fig. 2 and Fig. 3).

Furthermore, we identify several interlayer modes in the range between 0.5~1.0 THz. Specifically, the low-frequency peak observed at ~0.7 THz in the experiments (Figs. 2 and 3) is in good agreement with the interlayer shear mode frequencies at 0.73 THz for $T_A$ stacking and 0.68 THz and 0.75 THz for the two shear directions in $T_AT_C$ stacking. Note that in $T_A$ stacking, the two orthogonal interlayer shear modes are degenerate, whereas this degeneracy is lifted in $T_AT_C$ stacking because the two shear directions are no longer equivalent. Likewise, the opposite-phase and double-layer modes are degenerate in the $T_A$ stacking structure due to symmetry.

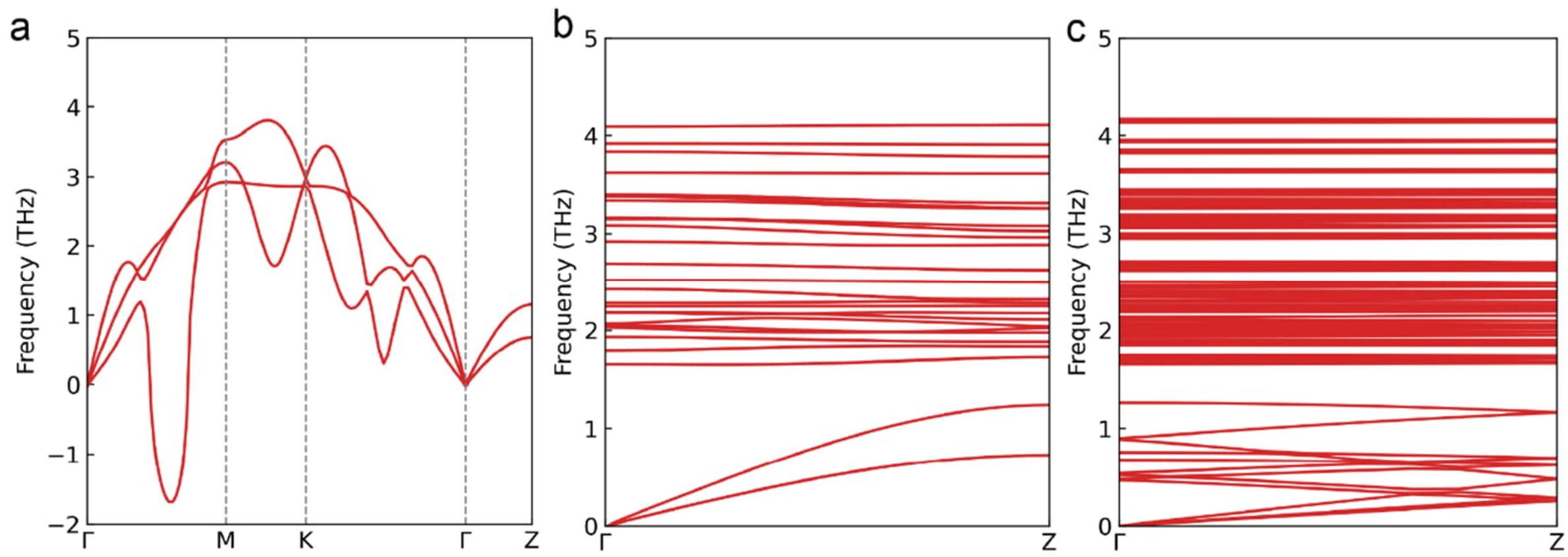

**Fig. S2: Low frequency phonon dispersion of 1T-$TaS_2$ in different phases. a** Undistorted 1T-$TaS_2$ with $T_A$ stacking. **b** CDW phase of 1T-$TaS_2$ with $T_A$ stacking. **c** CDW phase with $T_AT_C$ stacking.

| Mode frequency (THz) | Breathing | Opposite-phase breathing | Double-layer breathing | Shear | Opposite-phase shear | Double-layer shear |
|---|---|---|---|---|---|---|
| CDW $T_A$ | 1.24 | 0.84 | 0.84 | 0.73 | 0.51 | 0.51 |
| CDW $T_AT_C$ | 1.27 | 0.90 | 0.88 | 0.68/0.75 | 0.47/0.52 | 0.48/0.54 |

**Tab. S1: Interlayer vibrational mode frequencies of CDW phase $TaS_2$.** The corresponding vibrational modes are illustrated in Fig. S3 and S4.

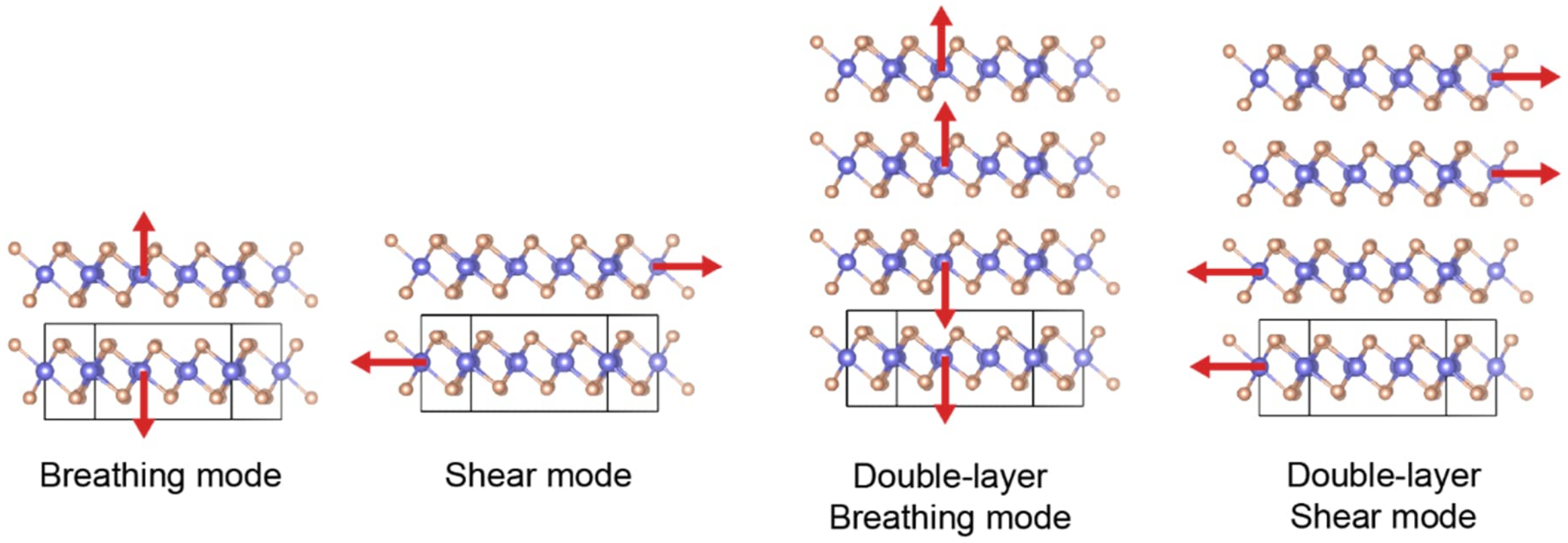

**Fig. S3: Sketch of interlayer vibration modes of $T_A$ stacking of the CDW phase of 1T-$TaS_2$.** Red arrows indicate the vibration directions. The corresponding frequencies are listed in table S1.

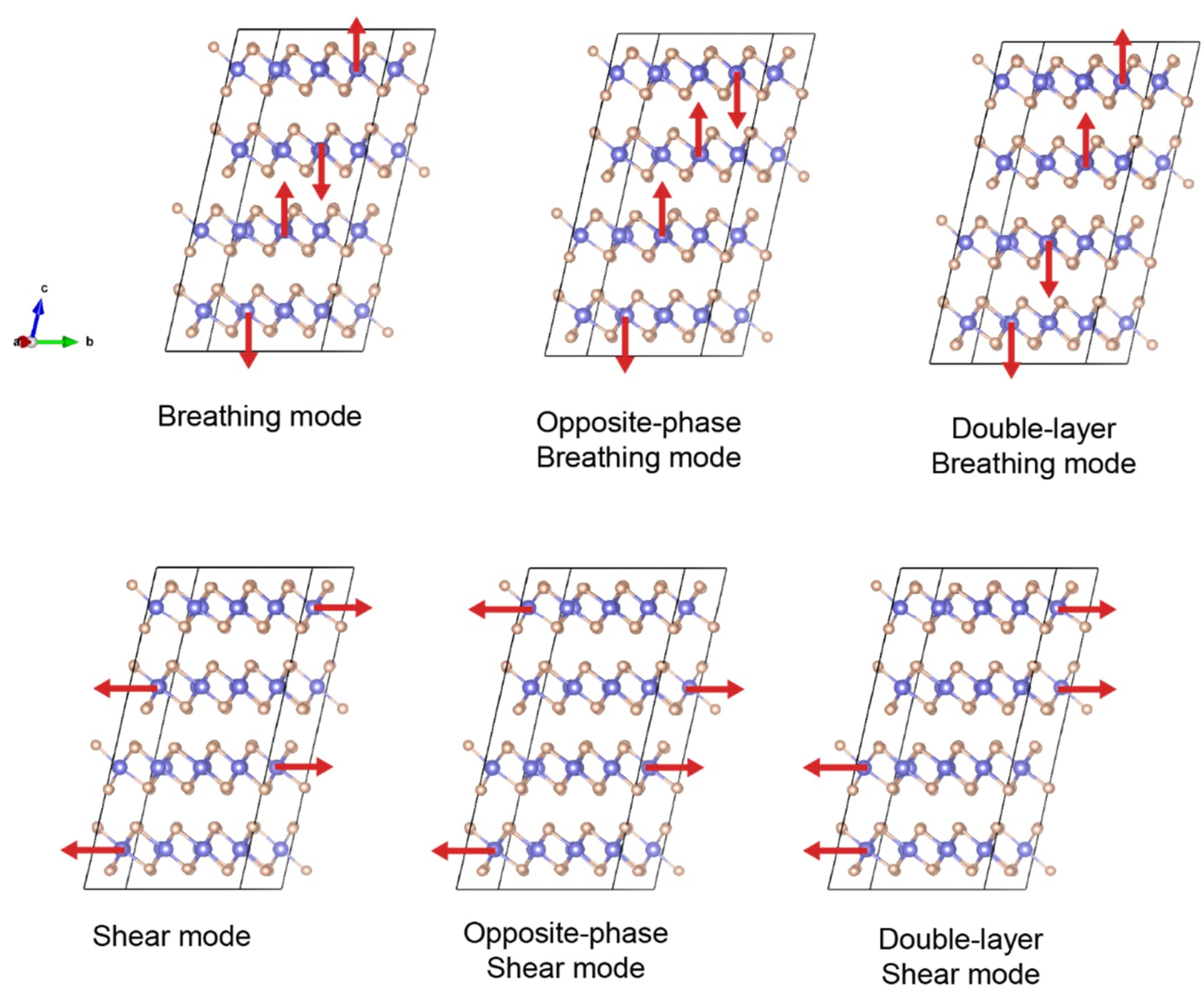

**Fig. S4: Sketch of interlayer vibration modes of $T_AT_C$ stacking of the CDW phase of 1T-$TaS_2$.** Red arrows indicate the vibration directions. The corresponding frequencies are listed in table S1.

## 3. Additional THz-STM images

Fig. S5 shows the topography and THz-STM images of the region in which the pink real-time oscilloscope trace in Fig. 5b in the main manuscript was recorded (pink circle position).

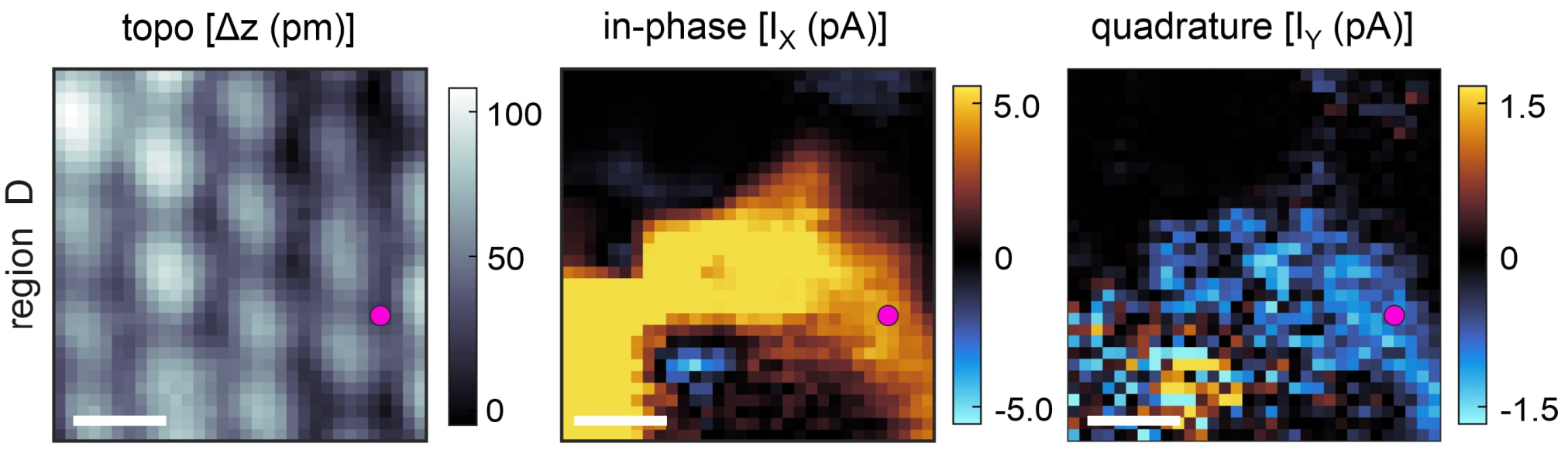

**Fig. S5: THz-induced metastability on another area of the sample.** Topographic (left) and THz-STM images of the region in which the oscilloscope trace in Fig.5b was recorded. The exact measurement position is marked by a pink circle. Both the in-phase ($I_X$, center) and quadrature components ($I_Y$, right) of the lock-in signal are plotted. The images were simultaneously acquired at a setpoint parameter of $V_{dc} = -100$ mV, $I_{dc} = 60$ pA, $E_{THz} = 390$ V/cm. Scalebar: 1 nm.

## 4. Simulation of lock-in detection of THz-induced currents

The observation of a finite signal $I_Y$ in THz-STM indicates that the chopped THz pulses modulate the current on timescales of the inverse chopper frequency $1/f_{chop}$. To illustrate this, we simulate the lock-in detection scheme in Fig. S6, where we assume a perfect square-wave modulation of the THz pulse train at $f_{\mathrm{chop}}$ and an infinite measurement bandwidth. If the tunnelling current $I(t)$ responds instantaneously to the modulation, as in THz-lightwave-driven STM, $I(t)$ follows the square-wave modulation and the lock-in phase remains zero ($\varphi_{\mathrm{lockin}} = 0$), i.e, the first harmonic S1 of the demodulated signal is in-phase with the reference (light blue curves in Fig. S6b). However, if the current responds delayed, the rising and/or falling edges of $I(t)$ will be temporally distorted (dark blue and pink curves in Fig. S6b,c , with different rise/fall times as example), resulting in $\varphi_{\mathrm{lockin}} \neq 0$ and a non-zero quadrature signal $I_{\mathrm{Y}}$. The time-averaged current $I_{\mathrm{dc}}$ and the magnitude of the lock-in signal $R_{\mathrm{lock-in}}$ depend on the rising and falling edges of the chopper half-cycles, i.e., the formation time and lifetime of the MS. In particular, for very slow dynamics with an incomplete signal recovery between two-half cycles, $R_{\mathrm{lock-in}}$ decreases while $I_{\mathrm{dc}}$ keeps increasing.

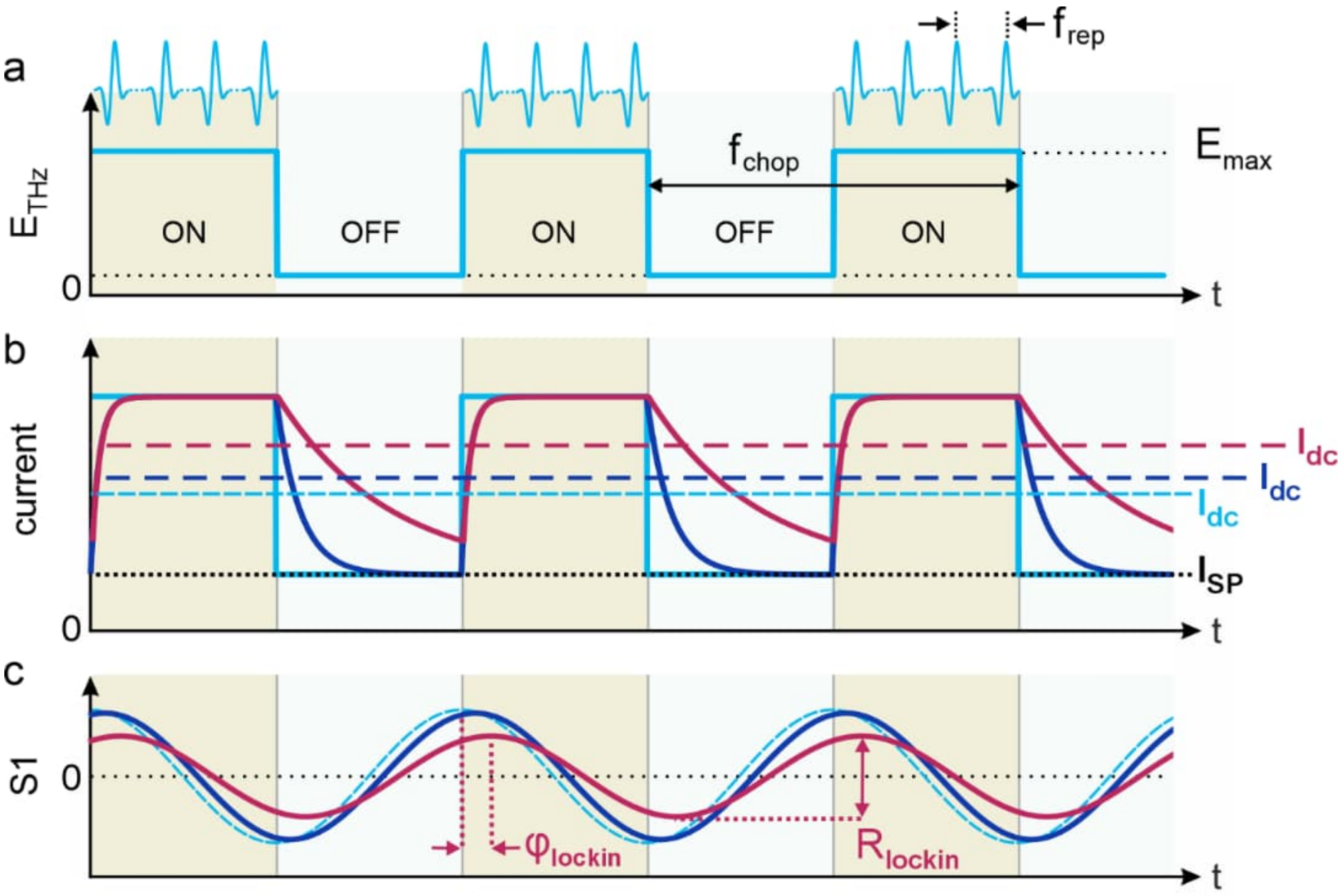

**Fig. S6: Simulation of lock-in detection of THz-induced currents. a** Square-wave modulation of the THz pulse train at the chopper frequency $f_{chop}$. **b** Resulting modulation of the tunnelling current $I(t)$ for an instantaneous (cyan) and delayed (blue and pink) response of $I(t)$. **c** Corresponding first harmonic $S_1$ of the demodulated signal. Delayed rising and/or falling edges lead to a lagging phase ($\varphi_{lockin} < 0$) and a signal $I_Y$ with opposite polarity to $I_X$. Furthermore, both the amplitude $R_{lockin}$ and the time-averaged current $I_{dc}$ (dashed lines in b) depend on the temporal profile of $I(t)$. $I_{SP}$ indicates the dark set point current. An infinite bandwidth of the electronics is assumed for simplicity.

### 5. Effect of STM preamplifier gain on real-time tunnelling response

Even when the tunnelling current responds quasi-instantaneously to the incident THz field (as in lightwave-driven THz-STM), the real-time response of the current will usually not directly follow the THz power modulation by the chopper, but will be low-pass filtered due to the finite bandwidth of the STM preamplifier. To illustrate this, we show in Fig. S7 the real-time tunnelling response for preamplifier gains of $10^8$ V/A and $10^9$ V/A, corresponding to 10 kHz and 1 kHz bandwidth, respectively. The 1 kHz bandwidth at the $10^9$ gain clearly distorts the real-time waveform of the tunnelling signal. This demonstrates the importance of using the correct reference signal when measuring a metastable response using the quadrature signal $I_Y$.

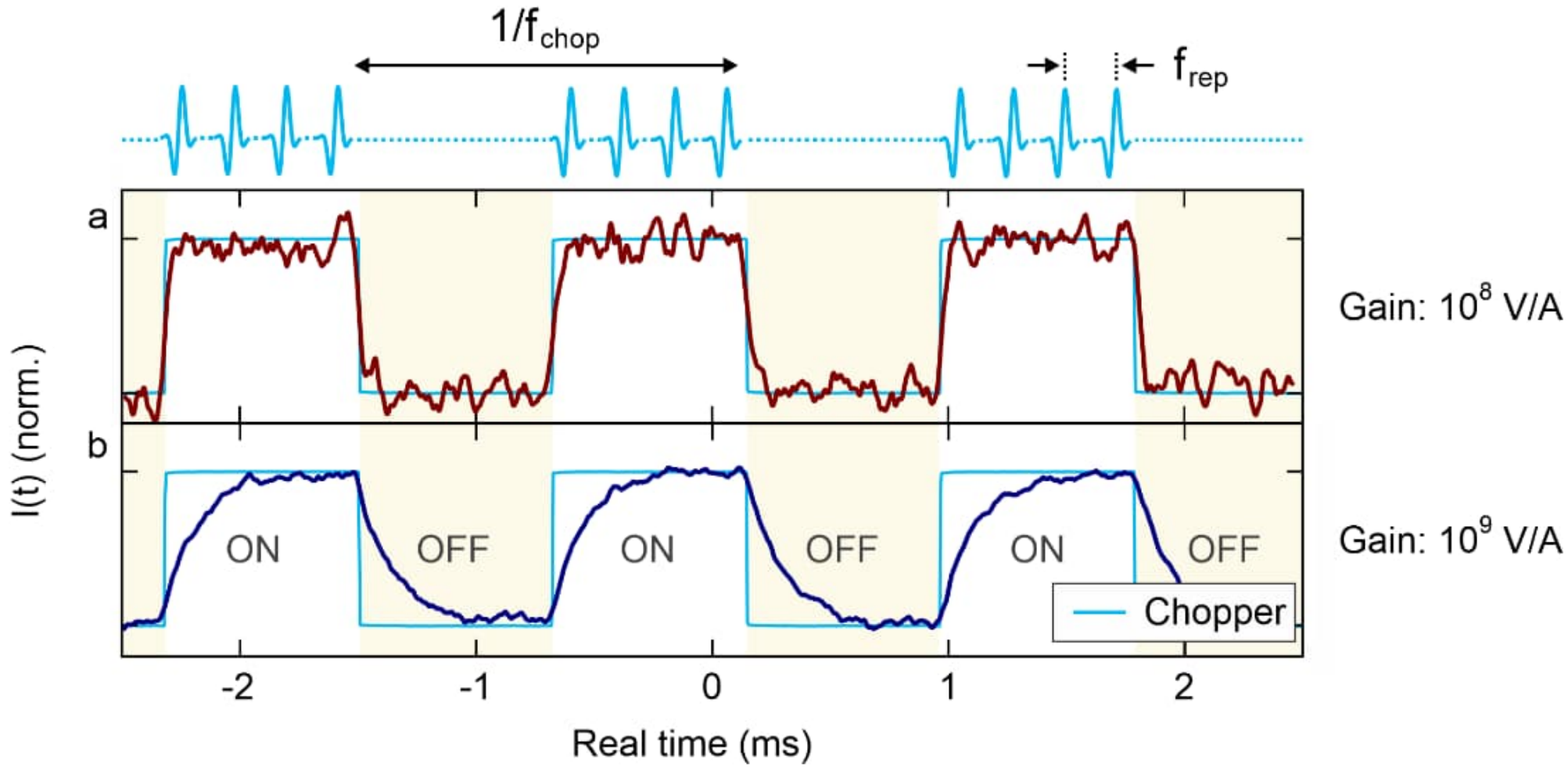

**Fig. S7: Real-time response of the STM Preamplifier. a,b** Real-time response of photoemission currents I(t) measured with the STM preamplifier using a gain of (a) $10^8$ V/A and (b) $10^9$ V/A. The light blue square indicates the modulation of the incident THz power by the chopper. Because the preamplifier bandwidth depends on its gain, it sets the timescale over which the photocurrent response can be measured. Since the photoemission current responds quasi-instantaneously to the THz field, these real-time traces reflect the gain-dependent instrument response function. Traces were recorded at $V_{\mathrm{dc}} = 6$ V, $E_{\mathrm{THz}} = 3.9$ kV/cm, $\phi_{NIR} = 4.5\ \mathrm{mJ/cm^2}$.

### 6. Ultrafast THz-STM in the presence of THz-induced metastability

We present several observations showing that ultrafast THz-lightwave-driven tunnelling can originate from a THz-induced metastable state, and that ultrafast dynamics and metastability can be observed simultaneously during THz-STM measurements. In particular, we discuss the extent to which both $\Delta I_{\mathrm{LW}}$ and $\Delta I_{\mathrm{LDOS}}$ contribute to the THz-driven current measured via lock-in detection in a single THz-STM measurement, and how to distinguish both experimentally. We explore this by taking a closer look at the data recorded in area 3, where the ultrafast dynamics in Fig. 3 are measured.

Figs. S8a,c show THz-STM images of the in-phase ($I_{\mathrm{X}}$) and quadrature ($I_{\mathrm{Y}}$) components of the lock-in signals  measured in area 3. Figs. S8b,d show the same images with a rescaled color scale. A significant signal $I_{\mathrm{Y}}$ is observed over the whole area. As discussed in Figs. 5 and

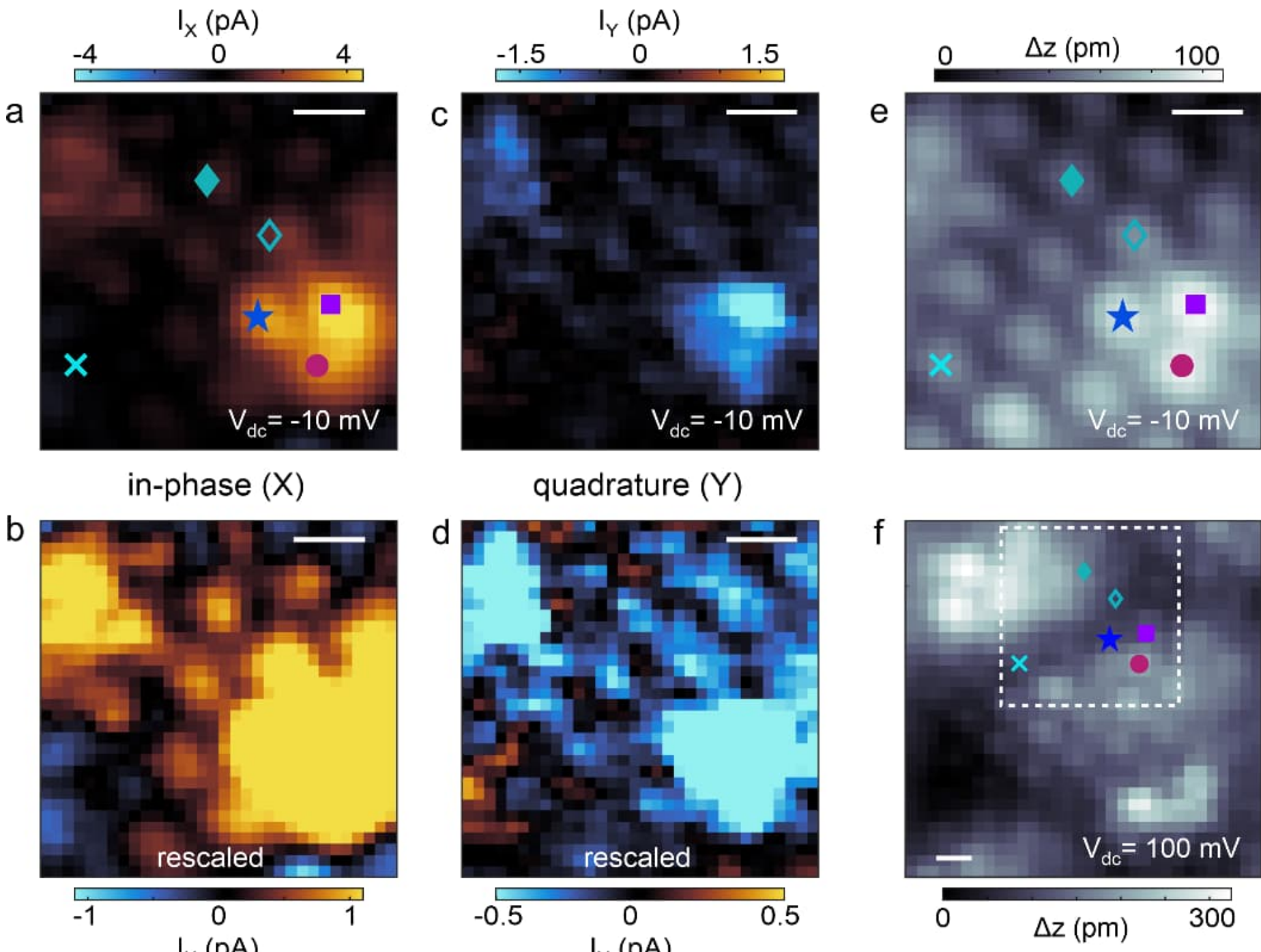

**Fig. S8: Metastable response in area 3. a-d** Constant-current THz-STM images of the in-phase ($I_{\mathrm{X}}$, a-b) and quadrature ($I_{\mathrm{Y}}$, c-d) components of the lock-in signal recorded on area 3. The images in (b) and (d) are rescaled from (a) and (c) and reveal the CDW pattern outside the defect. **e** Simultaneously recorded topographic image of the same area (same as in Fig. 3). **f** In-gap STM image of the region shown in Fig. 3a recorded at $V_{\mathrm{dc}}$ = 100 mV, $I_{\mathrm{dc}}$ = 10 pA, revealing a contrast indicative of a buried domain boundary. The blue star, pink circle and purple square show the defect positions of the ultrafast measurements shown in Fig. 3. The images a-e were recorded at setpoint parameters of $V_{\mathrm{dc}} = -10$ mV, $I_{\mathrm{dc}}$ = 50 pA and $E_{\mathrm{THz}}$ = 780 V/cm. Scale bar: 1 nm.

S6, this results from a metastable response on millisecond time scales. Fig. S8e shows the simultaneously recorded topographic STM image (same as Fig. 3b). Fig. S8f shows a larger-scale STM image recorded at an in-gap bias of $V_{\mathrm{dc}} = 100$ mV. In area 3 (white dashed box), a dark line indicative of a domain boundary is observed. This contrast is absent in STM images recorded at bias voltages outside the insulating gap (e.g. Figs. 3b/S8e). Since the LDOS at the 1T-$TaS_2$ surface is sensitive to subsurface stacking, this contrast likely originates from a buried CDW domain boundary below the surface. The colored symbols indicate where the STS spectra and ultrafast dynamics shown in Fig. 3 were recorded. Interestingly, these positions are located directly on top of a buried domain boundary. This finding supports our hypothesis that the STS modulations observed in Fig. 3c result from spatial variations and irregularities in the stacking of SoD clusters, and that the metastable state involves stacking changes which are excited during THz-STM operation.

We note that in most positions in area 3, the sign of $I_{\mathrm{X}}$ is opposite of what is expected for lightwave-driven THz-STM. Specifically, $I_{\mathrm{X}}$ is positive in most regions, indicating a net current flow from tip to sample. However, for the gapped STS spectra shown in Fig. 3c, rectification from the occupied LB states should yield a negative $I_{\mathrm{X}}$ in all these positions at the applied bias of $V_{\mathrm{dc}} = -10$ mV (see calculations in Fig. S14c in the appendix). In addition, the magnitude of $I_{\mathrm{X}}$ reaches much larger values than what is expected for lightwave-driven THz-STM at our conditions. These observations indicate that another mechanism, such as THz-induced LDOS changes, must contribute to the THz-driven current.

To further examine this, we measure in Figs. S9a,b the dependence of the time-averaged current $|I_{\mathrm{dc}}|$ (top) and $I_{\mathrm{X}}$ (bottom) on the incident THz peak field $E_{\mathrm{THz}}$. The curves are measured for two sample locations which yield an opposite sign in $I_{\mathrm{X}}$ (cyan cross and blue star in Fig. S8) and for two bias voltages $V_{\mathrm{dc}} = \pm 10$ mV. The data is recorded at constant tip height (details see figure caption). In both positions, $|I_{\mathrm{dc}}|$ and $I_{\mathrm{X}}$ change considerably above a position-dependent threshold field $E_{\mathrm{THz,th}}$. Moreover, the THz-induced change of the time-averaged current, $\Delta I_{\mathrm{dc}}$, can be of comparable (blue star) or much larger (cyan cross) magnitude than the simultaneously recorded $I_{\mathrm{X}}$. The latter scenario is not expected for lightwave-driven THz-STM, for which $\Delta I_{\mathrm{dc}} = I_{\mathrm{X}}$. Instead, this behavior ($\Delta I_{\mathrm{dc}} \gg I_{\mathrm{X}}$) is consistent with a THz-induced metastable change of the LDOS ($\Delta I_{\mathrm{LDOS}}$), whose lifetime significantly exceeds the inverse chopper frequency, such that the signal recovery between chopper half-cycles is incomplete (see $I_{\mathrm{dc}}$ for the pink curve in Fig. S6).

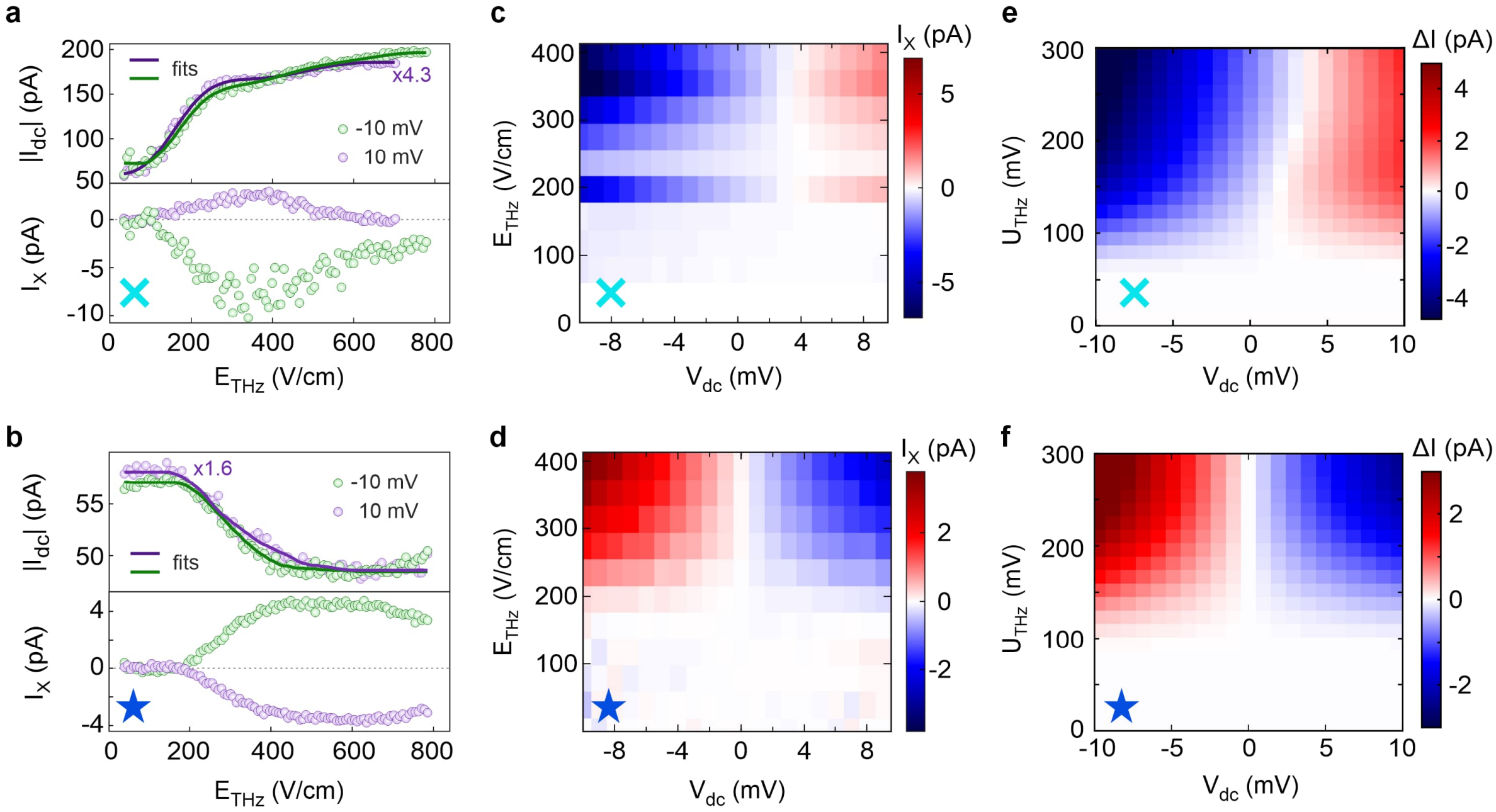

**Fig. S9: THz-induced tunnelling spectroscopy on 1T-TaS₂. a,b** Time-averaged current $|I_{\mathrm{dc}}|$ (top-panels) and THz-induced current $I_{\mathrm{X}}$ (bottom panels) as function of $E_{\mathrm{THz}}$ for $V_{\mathrm{dc}} = \pm 10$ mV at the positions marked by the cyan cross and blue in Fig. 3 and Fig. S8. Above a position-dependent threshold field, $|I_{\mathrm{dc}}|$ increases (decreases) at the cyan cross area (blue star area) with similar behavior for $V_{\mathrm{dc}} = \pm 10$ mV. $I_{\mathrm{X}}$ exhibits a comparable threshold and can differ significantly from $|I_{\mathrm{dc}}|$ (cyan cross, a) or sometimes closely follows $|I_{\mathrm{dc}}|$ (blue star, b). The solid lines are polynomial fits used for the simulations described in the appendix. **c,d** Measured $I_{\mathrm{X}}$ as a function of $E_{\mathrm{THz}}$ and $V_{\mathrm{dc}}$. While in (c), the polarity of $I_{\mathrm{X}}$ matches that of $V_{\mathrm{dc}}$, both have opposite sign in (d). In both cases, a sharp polarity inversion of $I_{\mathrm{X}}$ is observed near $V_{\mathrm{dc}} \approx 0$ mV. All data are recorded at constant tip height set at $V_{\mathrm{dc}} = -10$ mV and $I_{\mathrm{dc}} = 50$ pA without illumination. **e,f** Simulated total current $\Delta I$ including a THz-induced change of the LB peak as described in the appendix. The shifted $I_{\mathrm{X}} = 0$ contour that is clearly visible in (e) originates from contributions due to $\Delta I_{\mathrm{LW}}$ within the metastable state.

We further observe that the sign and magnitude of $I_{\mathrm{X}}$ exhibit a striking sensitivity to small millivolt-scale bias variations around $V_{\mathrm{dc}} = 0$ mV. Figs. S9c and S9d show maps of $I_{\mathrm{X}}$ as function of $V_{\mathrm{dc}}$ and $E_{\mathrm{THz}}$ for the same locations as in Fig. S9a and S9b, measured in the bias range $V_{\mathrm{dc}} = \pm 10$ mV. In both cases, the sign of $I_{\mathrm{X}}$ inverts within an energy window of less than $\pm 1$ mV near $V_{\mathrm{dc}} = 0$ mV. This behavior is characteristic of a quasi-stationary THz-induced modification of the LDOS,[8] observed here within a narrow energy window around $E_{\mathrm{F}}$. The abrupt sign inversion of $I_{\mathrm{X}}$ and its disappearance ($I_{\mathrm{X}} = 0$) near $V_{\mathrm{dc}} = 0$ mV occurs because no tunnelling is allowed at zero bias, regardless of possible changes in the LDOS at $E_{\mathrm{F}}$. The overall opposite sign of $I_{\mathrm{X}}$ for the two locations (cyan cross and blue star) shows that the THz-induced LDOS changes near $E_{\mathrm{F}}$ are opposite in both locations. Specifically, the fact that the same (opposite) polarity of $I_{\mathrm{X}}$ compared to $V_{\mathrm{dc}}$ is observed at the cyan cross (blue square) implies that the LDOS increases (decreases) near $E_F$ due to the THz excitation. This shows the pronounced spatial inhomogeneity of $\Delta I_{\mathrm{LDOS}}$ in our measurements.

In the measurements in Figs. S9c and S9d, the THz-induced currents $I_X$ are dominated by THz-induced LDOS changes ($\Delta I_{LDOS}$). However, the offset bias of ~3.5 mV at which $I_X$ changes sign in Fig. S9c (dashed line) can be explained by additional contributions from ultrafast rectification. We corroborate this by numerically simulating the $I_X - E_{THz} - V_{dc}$ maps in Fig 9 (details are explained in the appendix). For this purpose, we calculate $\Delta I = \Delta I_{LDOS} + \Delta I_{LW}$ including contributions from THz-lightwave-driven tunnelling ($\Delta I_{LW}$) and quasi-stationary THz-induced LDOS changes ($\Delta I_{LDOS}$) using the STS spectra in Fig. 3c. For the position with the "clean" STS resembling that of the $T_A$-stacked surface termination (cyan cross), we can reproduce the experimental $I_X - E_{THz} - V_{dc}$ maps assuming a THz-induced splitting of the LB peak (see Fig. S13), similar to the experimentally observed LB splitting in Fig. 4c. Conversely, at the position of the blue star, we fit the initially distorted LB peak by two Gaussians and can reproduce the experimental $I_X - E_{THz} - V_{dc}$ maps assuming a THz-driven coalescence of the two peaks (see Fig. S14). Similar behaviors are observed for the other two positions (purple square and pink circle) on top of the CDW irregularity.

Notably, although the exact contributions of $\Delta I_{LDOS}$ and $\Delta I_{LW}$ cannot be fully resolved experimentally, the ~3.5 mV offset observed in Fig. S9c is a clear indication for a contribution from $\Delta I_{LW}$. Due to the large separation of time scales, the rectified current $\Delta I_{LW}$ is hereby governed by the quasi-stationary LDOS of the THz-induced metastable state. Therefore, on the one hand, the THz pulses drive the system into a metastable state (THz as pump), while on the other hand, they serve as an ultrafast probe to measure the metastable state's ultrafast response to NIR excitation (THz as probe).

To illustrate this dual role, Fig. S10 shows a sketch of the THz pulse sequence within a single chopper "ON" cycle (compare Figs. 5 and S6). At $f_{chop} = 607$ Hz and $f_{rep} = 2$ MHz, roughly 1650 pulses arrive at the sample during one chopper "ON" half-cycle. In phase 1, directly after the chopper opens (gray area in Fig. S10a), the first arriving THz pulse(s) act as pump, driving the 1T-$TaS_2$ surface into a metastable stacking configuration. In phase 2 (yellow area), all the subsequent THz pulses can drive lightwave-rectification from this metastable state, allowing to probe its NIR-photoinduced femtosecond dynamics. Importantly, in order to be able to probe coherent dynamics while averaging over many pulse pairs, every NIR-THz pulse pair must probe the exact same ultrafast sample response. Therefore, coherent modes can only be measured reliably if phase 1 is short (or absent, as in conventional THz-STM) compared to phase 2. In this case, the metastable state is created quickly and persists sufficiently long so that the majority of THz pulses can probe ultrafast dynamics within that state. We note that we are not always able to detect an ultrafast coherent response in our measurements. We attribute this to situations in which the aforementioned condition is not fulfilled, but where successive pulse pairs probe different sample conditions. In this case, the average over non-identical material responses prevents us from observing coherent mode dynamics.

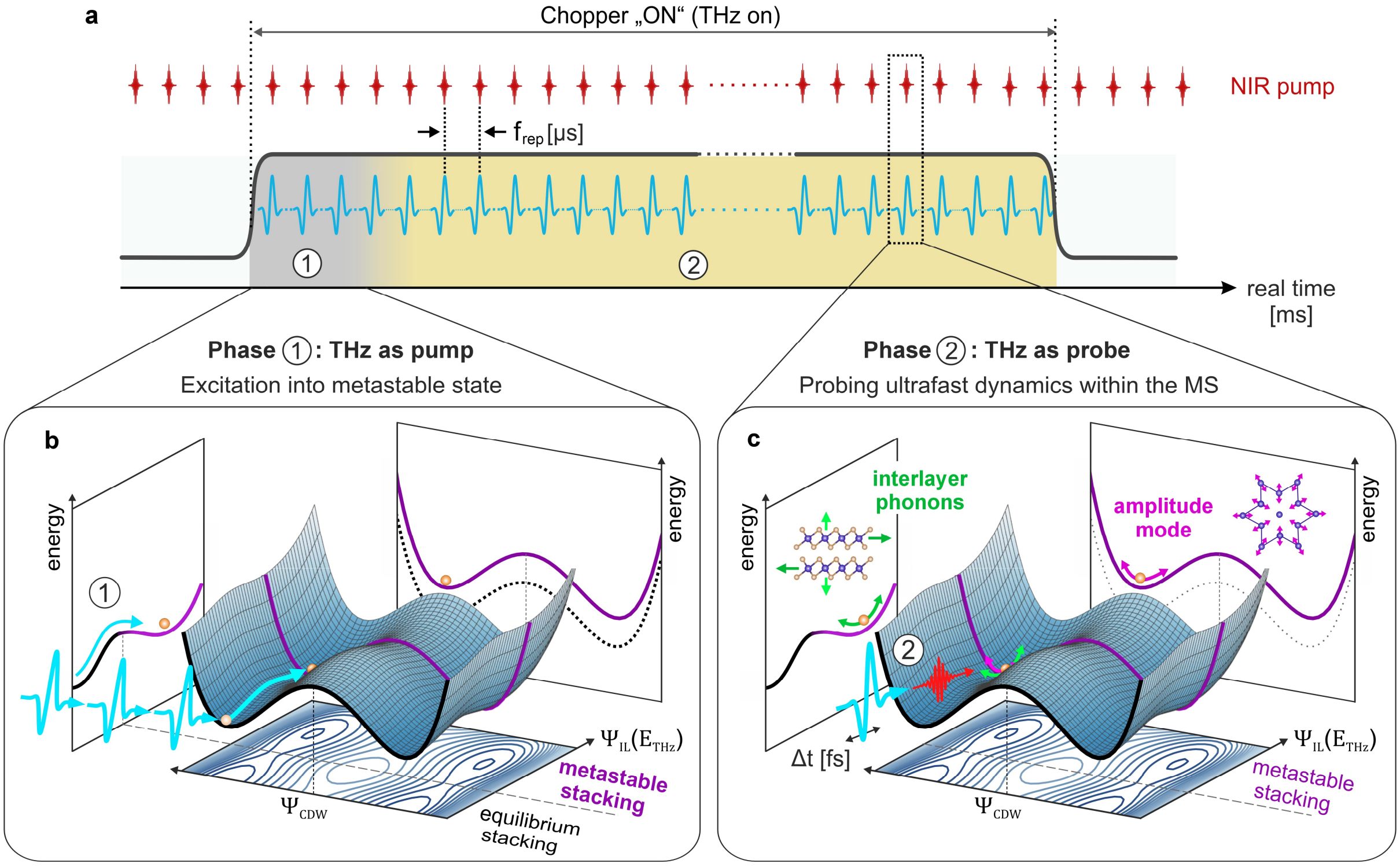

**Fig. S10: Dual role of THz excitation and overview of timescales. a** Illustration of the THz pulse train within one chopper half-cycle, during which >1500 THz pulses arrive at the sample. **b** The first arriving one or few THz pulse(s) act as pump and drive the $TaS_2$ surface into a long-lived metastable state (phase 1, gray area in a). **c** Each of the many subsequent THz pulses acts as probe for measuring ultrafast dynamics within the metastable state induced by femtosecond NIR pulses (phase 2, yellow area in a). In order to measure coherent phonon dynamics, the initially THz-induced metastable state must persist and remain stationary for the remainder of the chopper half-cycle period.

Finally, we note that the THz-induced changes to $I_{\mathrm{dc}}$ observed in Fig. S9a,b are fully reversible, and that no persistent changes are observed after THz illumination at the applied THz fields. Furthermore, although sample heating and increased carrier screening due to NIR photoexcitation will likely affect the generation and lifetime of the metastable state as well as the relative contributions of $\Delta I_{\mathrm{LDOS}}$ and $\Delta I_{\mathrm{LW}}$, we emphasize that THz-induced metastability persists under NIR excitation. This is evident from the same unconventional polarity behavior and sensitivity of $I_{\mathrm{X}}$ to $V_{\mathrm{dc}}$ under NIR illumination and from the delay-independence of $I_{\mathrm{Y}}$.

## 7. Discussion of possible mechanisms for coherent phonon detection

We briefly discuss the possible mechanisms by which the observed phonon modes may coherently modulate the tunnelling current on ultrafast time scales. In general, any coherent modulation of the LDOS underneath the tip by the excited phonons will modulate the tunnelling current. In the simplest scenario, this could arise from a geometrical gap size modulation, i.e. an out-of-plane motion of surface atoms underneath the tip.[9,10] However, in a material like 1T-$TaS_2$ with strong electron-lattice coupling, small structural distortions can lead to significant modulations of the electronic states and the insulating gap in the LDOS. In 1T-$TaS_2$, we expect such coupling to be pronounced not only for the intralayer AM, but also for the interlayer shear and breathing modes.

The AM phonon corresponds to a breathing motion of the SoD clusters (Fig. 3f), which modulates the intralayer CDW order and thus the insulating gap, as also shown by trARPES.[11–13] To study the effect of the AM on the LDOS, in Fig. S11a-b, we measure the dependence of $I_X$ on $E_{THz}$ for different NIR-THz delays over half an oscillation cycle of the AM (yellow box in Fig. S11c). The black dashed lines highlight constant $I_X$ values. Fig. S11b shows horizontal cuts at $\Delta t$ = 1.4 ps and $\Delta t$ = 1.6 ps, revealing a shift of the $I_X - E_{THz}$ curves by ~21 V/cm between opposite AM oscillation phases. We attribute this shift to an ultrafast coherent modulation of the insulating gap, consistent with coherent band shifts observed by time-resolved ARPES.[11,13,14] In contrast, a pure geometric gap size modulation would primarily rescale the tunnelling curve in Fig S11b rather than causing a shift along $E_{THz}$. Although a small out-of-plane motion of near-center sulfur atoms occurs during AM oscillation, we expect its effect on the tip-sample distance to be negligible at our experimental conditions. However, contributions from periodic gap size modulations cannot be fully excluded.

The interlayer breathing mode at 1.3 THz directly modulates the interlayer spacing and thereby the effective tip-sample distance. At the same time, it will coherently modulate the interlayer orbital overlap and interlayer hopping, leading to a periodic modulation of both hybridization-driven (band-insulating) and correlation-driven (Mott-insulating) contributions to the electronic gap. We therefore expect the breathing mode to affect both the geometric tip-sample distance and the LDOS within the insulating gap.

In contrast, the interlayer shear mode observed at ~0.7 THz will cause a lateral displacement of layers, which is unlikely to significantly affect the tip-sample distance. Instead, it will modulate the interlayer arrangement and lateral overlap of the vertically stacked SoD orbitals. Considering the strong dependence of the low-energy LDOS on the surface stacking,[15] we expect the shear mode to primarily redistribute spectral weight in the states near $E_{\mathrm{F}}$ that define the insulating gap. The enhanced shear mode amplitude observed locally at a single defect-like CDW cluster in Fig. 3 supports this interpretation, suggesting that shear-mode-induced LDOS modulations are particularly sensitive to local stacking irregularities and defects.

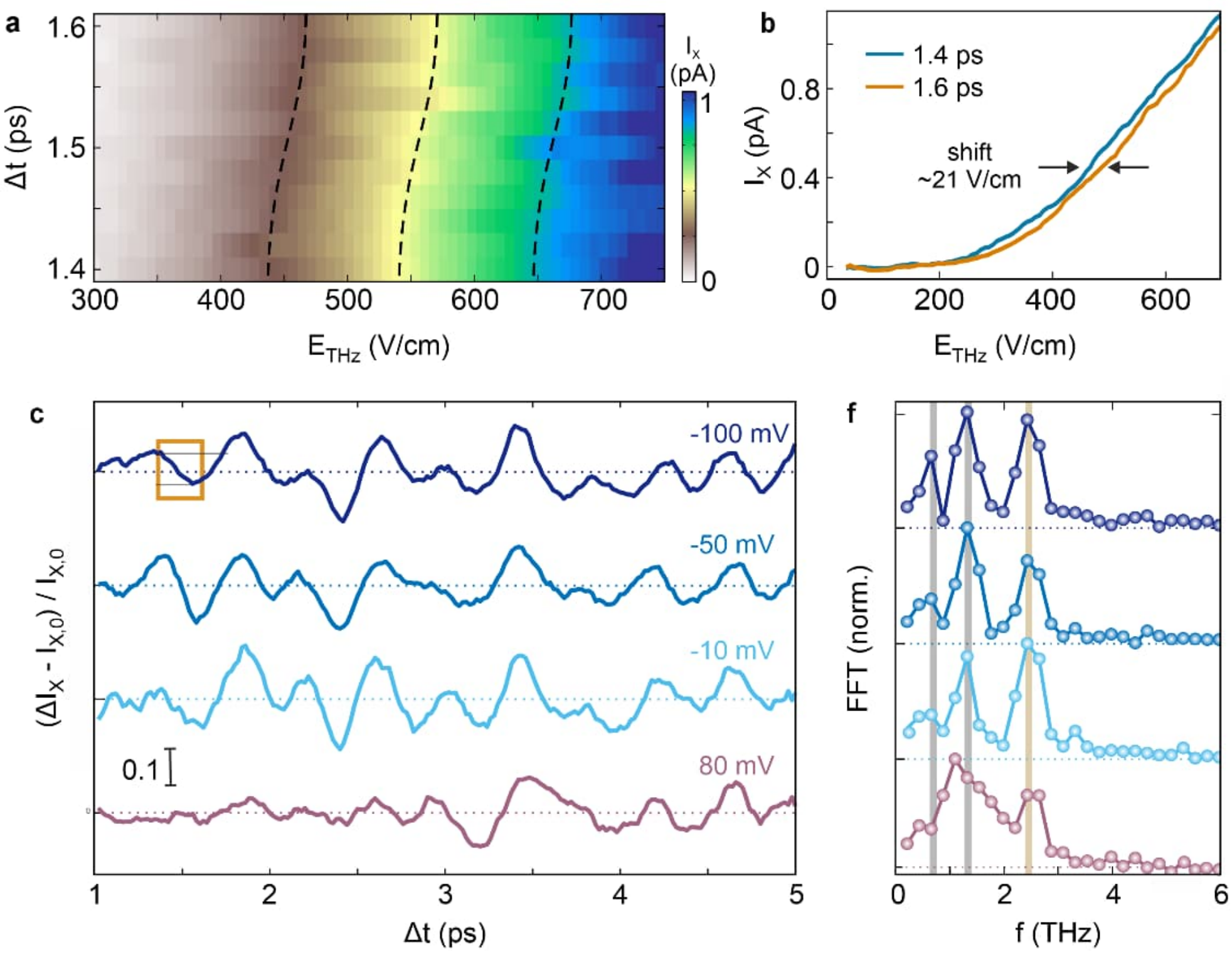

**Fig. S11. Phonon-phase dependent tunnelling. a** $I_{\mathrm{X}} - E_{\mathrm{THz}}$ curves measured as function of $\Delta t$ during a half oscillation cycle of the AM (yellow box in c). The black dashed lines are contours of constant $I_{\mathrm{X}}$. **b** Horizontal cuts $I_{\mathrm{X}}(E_{\mathrm{THz}})$ at $\Delta t = 1.4$ ps and $\Delta t = 1.6$ ps, revealing a shift of ~ 21 V/cm. **c** Phonon dynamics measured for different $V_{\mathrm{dc}}$. The THz-driven current is normalized to the non-oscillatory baseline $I_{\mathrm{X,0}}$ at 5 ps (see SI section 1). **d** Fourier-transform spectra of the data in (c).

An ultrafast modulation of the LDOS of the insulating gap is also consistent with the different oscillation patterns and FFT spectra we observe at a positive STM bias of 80 mV (Fig. S11c,d), which is in the middle of the insulating gap and thus energetically far away from any band onsets. In this case, the amplitude of the breathing mode is enhanced over that of the CDW amplitude mode and the shear mode. While more detailed time-resolved THz-STS will be required to gain further insight, this indicates that the breathing mode modulates the tip-sample distance more effectively, as expected. Further measurements and modelling are required to improve our understanding of the role of electronic and lattice degrees of freedom in measuring the ultrafast lattice dynamics of 1T-$TaS_2$ using THz-STM.

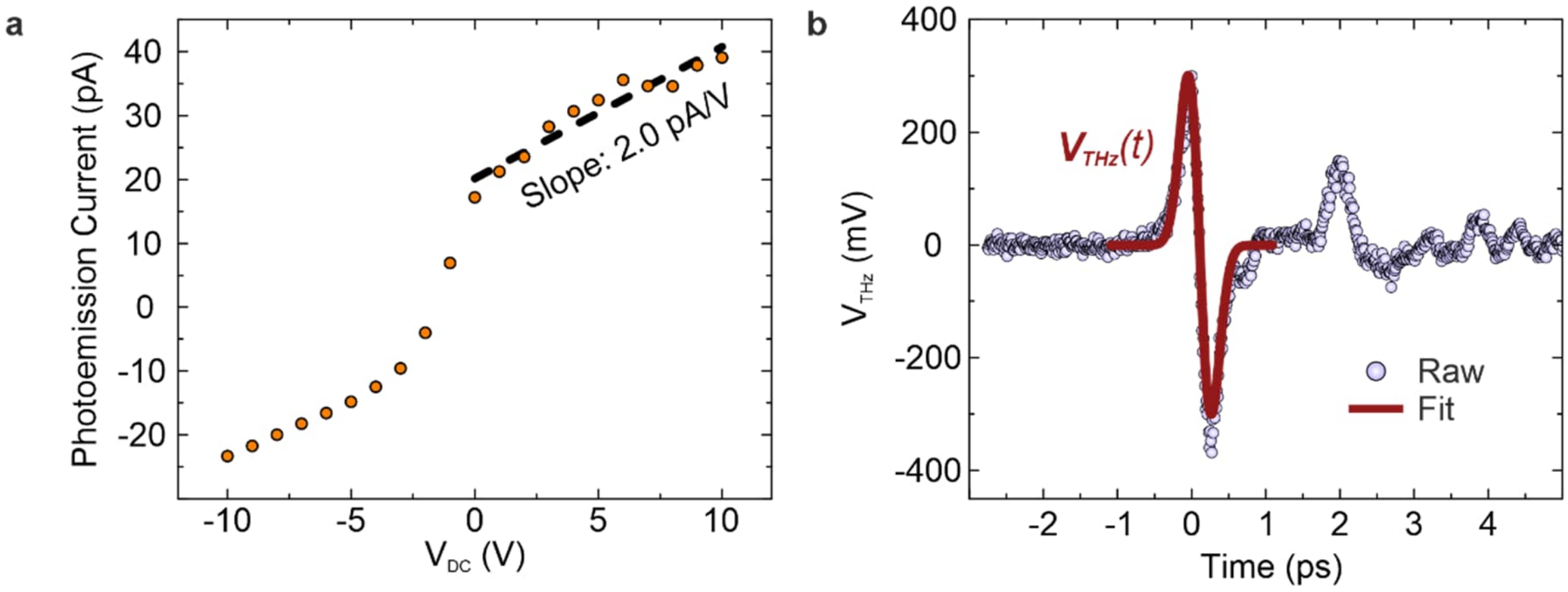

**Fig. S12. Photoemission sampling on Ag(111). a** Photoemission current ($I_{PES}$) as a function of the applied bias voltage. The THz-voltage is retrieved from a linear fit of the $I_{PES}$-$V_{dc}$ characteristic (black dashed line). **b** Tip-enhanced THz-waveform sampled at $V_{dc}$ = 6 V and $E_{THz}$ = 390 V/cm. The red line shows a fit of the waveform, which is used as input for the simulations. The NIR fluence was $\phi_{NIR}$ = 3.6 mJ/cm$^2$. Note that the time axis in b is inverted with respect to the NIR-pump – THz-probe traces shown in Fig. 2, 3 and S1, so that NIR-induced dynamics would appear on the left side of the THz waveform in this representation.

## 8. Photoemission sampling of THz bias transient

## APPENDIX: Simulation of THz-induced currents

The THz-driven tunnelling current is calculated using a one-dimensional Bardeen model at 10 K assuming a constant tip LDOS $\rho_t = 1$. In this case, the tunnelling current $I(V)$ is given by

$$I(V) = \gamma \int_0^{eV} \rho_s(\epsilon - eV) T(\epsilon, V, z) d\epsilon \qquad (1)$$

where $\epsilon$ is energy, $z$ is the tip-sample distance, $V$ is the voltage applied between tip and sample, $\rho_s(\epsilon - eV)$ is the sample LDOS, and $T(\epsilon, V, z)$ is the transmission probability which we approximate as a trapezoidal barrier.[16] The constant $\gamma$ is a free scaling parameter to adjust the magnitude of the simulated current to best match the measured I(V) characteristics.

To account for the dual effect of the THz pulses in inducing a femtosecond bias as well as quasi-stationary changes in the LDOS, the THz field enters the tunnelling model in two fundamentally different ways. First, to model THz-lightwave-driven tunnelling, we assume a time-dependent voltage $V(t) = V_{\text{dc}} + V_{\text{THz}}(t)$, where

$$V_{\text{THz}}(t) = V_{\text{THz,pk}} \exp\left(\frac{-t^2}{2{\sigma_{\text{THz}}}^2}\right) \sin(2\pi\omega_{\text{THz}} t) \qquad (2)$$

models the THz bias pulse with peak amplitude $V_{\text{THz,pk}}$, temporal width $\sigma_{\text{THz}}$ and center frequency $\omega_{\mathrm{THz}}$. We generate $V_{\mathrm{THz}}(t)$ by fitting eq. (2) to the measured tip-enhanced THz waveform (Fig. S12b), where $V_{\mathrm{THz,pk}}$ scales linear with the incident THz field $E_{\mathrm{THz}}$. The resulting ultrafast current $I(V(t), t)$ gives rise to a rectified tunnelling current

$$\Delta I_{\mathrm{LW}} = f_{\mathrm{rep}} \int_{-\infty}^{\infty} [I(V(t), t, \rho_{\mathrm{s}}) - I(V_{\mathrm{dc}}, \rho_{\mathrm{s}})] dt'. \quad (3)$$

The difference $[I(V(t), t) - I(V_{\mathrm{dc}})]$ accounts for the THz-induced changes to the current, so that $\Delta I_{\mathrm{LW}}$ yields the THz-lightwave-driven current as measured by lock-in detection.

Second, the THz-induced current $\Delta I_{\mathrm{LDOS}}$ caused by quasi-stationary changes of the LDOS is calculated by the difference

$$\Delta I_{\mathrm{LDOS}} = I_{\mathrm{THz,on}}(\rho_{\mathrm{s,MS}}) - I_{\mathrm{THz,off}}(\rho_{\mathrm{s,eq}}) \; . \quad (4)$$

Here, $I_{\mathrm{THz,on}}$ and $I_{\mathrm{THz,off}}$ are the tunnelling currents with and without THz illumination calculated by eq. (1) and $\rho_{\mathrm{s,MS}}$ and $\rho_{\mathrm{s,eq}}$ are the metastable and equilibrium LDOS, respectively. Equation (4) assumes that the lock-in detection captures the effective conductance difference between the equilibrium and long-lived metastable state. In reality, the temporal dynamics of the formation and decay of the metastable state as well as the detection bandwidth may also play a role. The total THz-induced tunnelling current $\Delta I$ is then the sum of both contributions,

$$\Delta I = \alpha \cdot \Delta I_{\mathrm{LDOS}} + \beta \cdot \Delta I_{\mathrm{LW}}, \quad (5)$$

where $\Delta I_{\mathrm{LDOS}}$ is calculated from eq. (4) and

$$\Delta I_{\mathrm{LW}} = f_{\mathrm{rep}} \cdot \int_{-\infty}^{\infty} \left[I\left(V(t), t, \rho_{\mathrm{s,MS}}\right) - I\left(V_{\mathrm{dc}}, \rho_{\mathrm{s,MS}}\right)\right] dt' \quad (6)$$

is now calculated from the $I$-$V$-characteristic of the metastable state determined by $\rho_{\mathrm{s,MS}}$. This is justified for the conditions discussed in Fig. S10, where the majority of THz pulses drive rectification from the metastable state. The assumption of a constant quasi-stationary LDOS $\rho_{\mathrm{s,MS}}$ is further justified by the clear separation of timescales between the 1 ps duration of the THz bias pulse and the millisecond lifetime of the metastable state. Since our model does not account for the actual dynamics of the metastable state, we note that $\Delta I$ in eq. (5) cannot be quantitatively compared to the lock-in amplitude measured in the experiment. The free parameters $\alpha$ and $\beta$ are therefore used to adjust the relative contributions of the currents to fit the magnitude of the measured lock-in signal $I_{\mathrm{X}}$. Specifically, $\alpha$ accounts for the fact that the lock-in amplitude depends on the actual time-dependence of $I(t)$ as shown in Fig. S6, and that the detailed LDOS changes are not known. The parameter $\beta$ accounts for the fact that details of the rectification process, such as the exact tunnelling conductance (metastable I-V-characteristic) and the precise THz waveform shape in tunnelling conditions, are not accurately known.

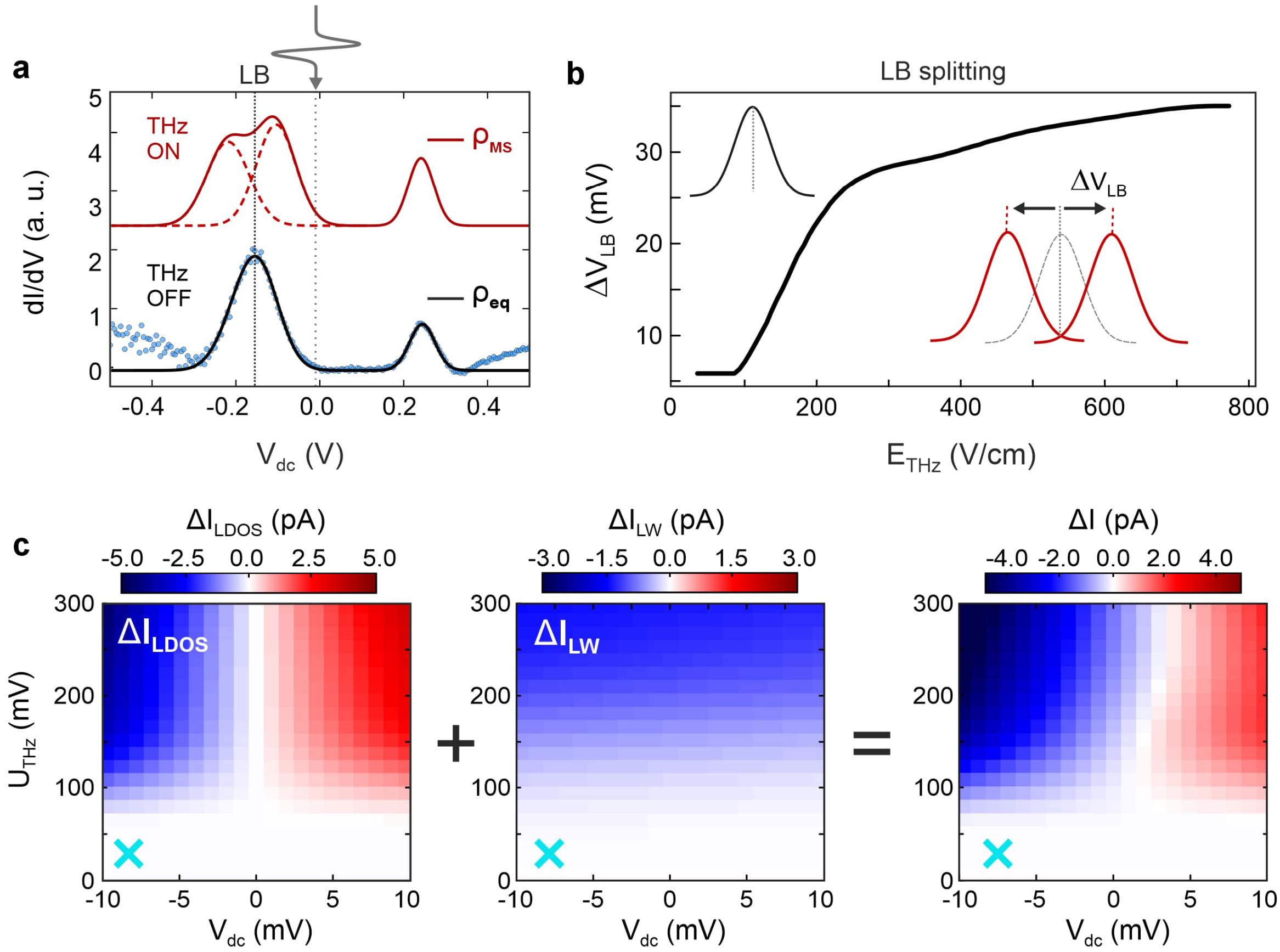

**Fig. S13. Simulation of THz-induced currents: Example 1. a** Measured STS (blue circles) on the cyan cross position in Fig. 3 and Fig. S8. The equilibrium LDOS ($\rho_{\mathrm{eq}}$, black line) was modeled by fitting the measured STS. The metastable state LDOS ($\rho_{\mathrm{MS}}$, red line) was calculated by symmetrically splitting the LB of $\rho_{\mathrm{eq}}$. The individual contributions to the metastable LB are shown as red-dashed lines. **b** Calculated LB band splitting magnitude as a function of $E_{\mathrm{THz}}$ extracted from the fit in Fig. S9a. **c** Simulated $\Delta \mathrm{I}_{\mathrm{LDOS}}$ (left) and $\Delta \mathrm{I}_{\mathrm{LW}}$ (center) as a function of $V_{\mathrm{THz}}$ and $V_{\mathrm{dc}}$. Their sum weighted by $\alpha = 0.056$ and $\beta = 340$ yields $\Delta I$ as shown in the right panel.

The equilibrium LDOS, $\rho_{\mathrm{s,eq}}$, used to calculate $\Delta I_{\mathrm{LDOS}}$ is obtained by fitting the measured STS spectra at a given sample position by a sum of Gaussian functions,

$$\rho_{\mathrm{s,eq}}(\epsilon) = \sum_{i=1}^{N} \mathrm{A}_{\mathrm{eq,i}} \exp\left(\frac{-\left(\epsilon - E_{\mathrm{eq,i}}\right)^2}{2\sigma_{\mathrm{eq,i}}^2}\right), \tag{7}$$

where $E_{\mathrm{s,i}}$, $A_{\mathrm{eq,i}}$ and $\sigma_{\mathrm{eq,i}}$ are the center position, amplitude and width of the Gaussian peaks.

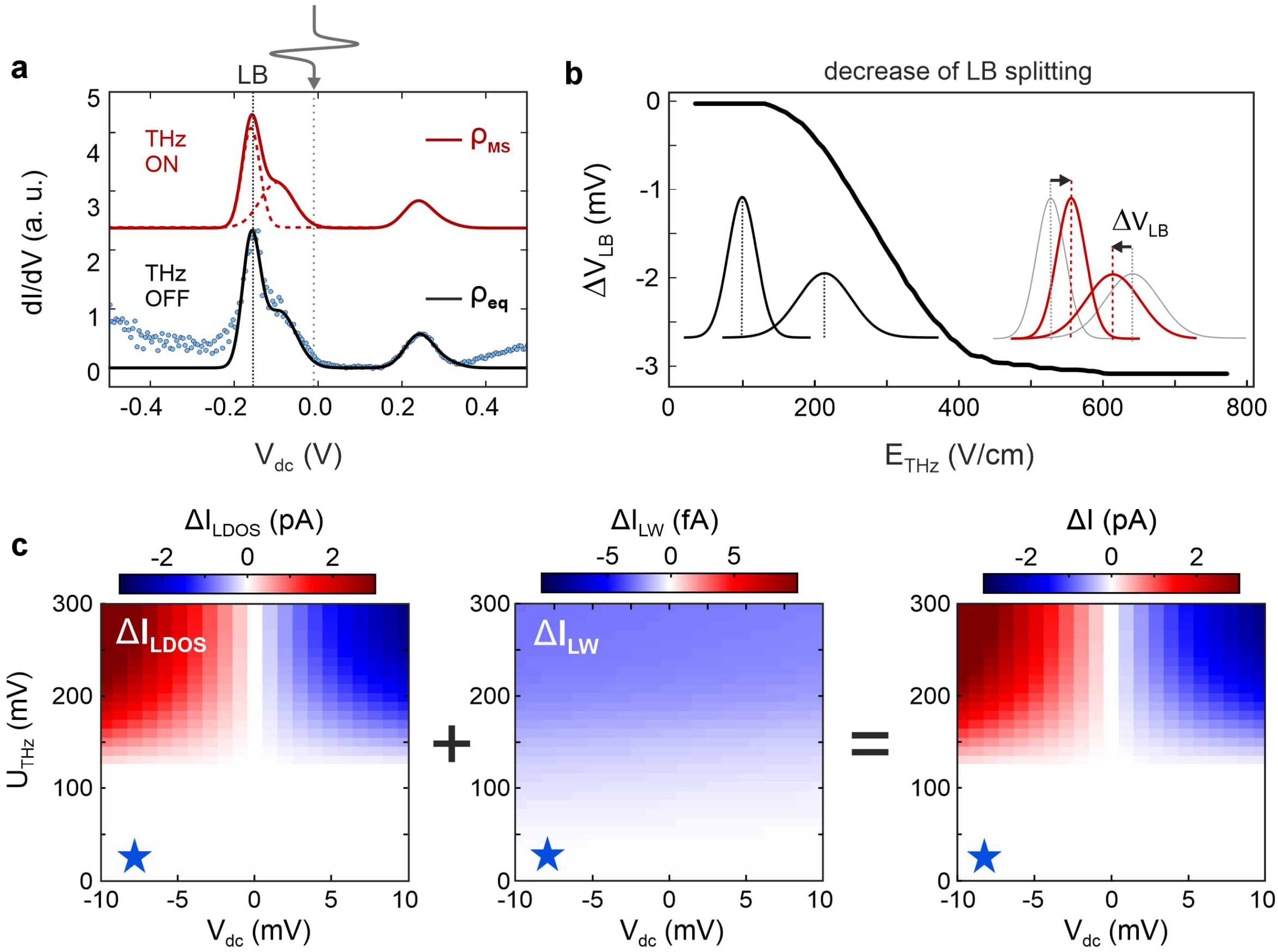

**Fig. S14. Simulation of THz-induced currents: Example 2. a** Measured STS (blue circles) on the blue star position in Fig. 3 and Fig. S8. The equilibrium LDOS ($\rho_{eq}$, black line) is modeled by fitting three peaks to the measured STS. The metastable state LDOS ($\rho_{MS}$, red line) is calculated by symmetrically shifting the two peaks forming the LB of $\rho_{eq}$. Their position in the metastable LB are shown as red-dashed lines. The shift of a few mV is too small to be clearly resolved visually here. **b** Calculated decrease of the LB band splitting as a function of $E_{THz}$ extracted from the fit in Fig. S9b. **c** Simulated $\Delta I_{LDOS}$ (left) and $\Delta I_{LW}$ (center) as a function of $V_{THz}$ and $V_{dc}$. Their sum weighted by $\alpha = 0.535$ and $\beta = 0.2$ yields $\Delta I$ as shown in the right panel.

Fig. S13a shows a fit of $\rho_{s,eq}(\epsilon)$ to the STS spectra measured on the position marked by a cyan cross in Fig. S8. We model the metastable LDOS, $\rho_{s,MS}$, assuming a THz-induced LB splitting similar to the experimental observations in Fig. 4d, reflecting the assumption that the THz field modifies the interlayer stacking and thus the width of the insulating gap in 1T-TaS$_2$.[17–20] For this purpose, we fit the LB peak by two Gaussians and symmetrically shift them by an amount $\Delta V_{LB}$ that is chosen such that the resulting calculated current follows the experimentally measured $I_{dc} - E_{THz}$ dependence (see fitted curve in Fig. S9a). The dependence of the splitting $\Delta V_{LB}$ on $E_{THz}$ is shown in Fig. S13b. For each combination of $V_{dc}$ and $E_{THz}$, we then calculate $\Delta I_{LDOS}$ using eq. (4). $\Delta I_{LW}$ is calculated using eq. (6) and by varying $V_{THz,pk}$ from 0 mV to 300 mV as estimated from photoemission sampling. In reality, partial field penetration into the sample will cause a THz voltage drop within the near-surface region resulting in somewhat lower THz fields. We note, however, that the simulation results remain qualitatively robust across a reasonable range of $V_{\mathrm{THz,pk}}$. The final $\Delta I$ is computed using eq. (5) with the free parameters $\alpha$ and $\beta$ described in the caption of S13. The resulting maps of $\Delta I_{\mathrm{LDOS}}$, $\Delta I_{\mathrm{LW}}$ and $\Delta I$ are shown in Fig. S13c. The THz-induced LB splitting, which effectively leads to an increase of the LDOS near $E_{\mathrm{F}}$, allows to reproduce well the experimental behaviour observed in Fig. S9c.

Fig. S14 shows similar calculations for the position of the blue star (Fig. S8), which exhibits the opposite polarity behavior compared to the data in Fig. S13. Here, $\rho_{\mathrm{s,eq}}(\epsilon)$ is obtained by fitting the distorted LB band in the equilibrium STS spectra by two Gaussians, which move closer together due to THz excitation ($\Delta V_{\mathrm{LB}} < 0$) rather than split further apart. In contrast to Fig. S13, this results in a decrease of the LDOS near $E_{\mathrm{F}}$. The dependence of $\Delta V_{\mathrm{LB}}$ on $E_{\mathrm{THz}}$ (Fig. S14b) is again obtained from the experimentally measured $I_{\mathrm{dc}} - E_{\mathrm{THz}}$ curve measured at the respective location (Fig. S9b). The calculated maps of $\Delta I_{\mathrm{LDOS}}$, $\Delta I_{\mathrm{LW}}$ and $\Delta I$ are shown in Fig. S14c, reproducing well the experimental map in Fig. S9d.

We note that also other THz-induced changes, such as a shift of the LB to higher (lower) energy or a closing (opening) of the insulating gap, would yield similar results as those in Fig. S13 (Fig. S14). To distinguish between the different possible scenarios, the LDOS changes would need to be explored in a larger bias range further away from $E_{\mathrm{F}}$ in more detail. A more quantitative determination of the THz-induced LDOS changes and the contributions from $\Delta I_{\mathrm{LW}}$ (i.e. a reliable determination of the parameters $\alpha$ and $\beta$) requires more comprehensive measurements and modelling, which is beyond the scope of the present work.